\documentclass[11pt,a4paper]{article}
\usepackage{bm,jcappub,Macro,tabularx}
\usepackage{url}
\usepackage{hyperref}
\usepackage{url}
\usepackage{epsf}
\usepackage{epsfig}
\usepackage{graphicx}
\usepackage{epstopdf}
\usepackage{enumitem}
\usepackage{tensor}	
\usepackage{multirow}
\usepackage{colortbl}
\usepackage{color}
\def\la{\langle}
\def\ra{\rangle}
\def\n{\noindent}
\def\be{\begin{equation}}
\def\ee{\end{equation}}
\def\ben{\begin{eqnarray}}
\def\een{\end{eqnarray}}
\def\nn{\nonumber}
\def\oh{\hat\Omega}
\def\myC{{\cal C}}

\def\myf{\Theta}

\def\br{{\bf r}}

\def\myC{{\cal C}}

\def\2p{{(2\pi)^2}}

\def\be{\begin{equation}}
\def\ee{\end{equation}}
\def\beq{\begin{equation}}
\def\eeq{\end{equation}}
\def\ben{\begin{eqnarray}}
\def\een{\end{eqnarray}}

\def\oh{{\hat\Omega}}
\def\nn{{\nonumber}}

\newcommand{\beqa}{\begin{eqnarray}}
\newcommand{\eeqa}{\end{eqnarray}}

\def\ikap0{{\cal J}_{\theta_0}(r)}

\def\one1{\langle \kappa_{(i)}\kappa_{(j)} \rangle}
\def\one{{[\bar \xi^{(ij)}]}}

\def\fsky{{f^{-1}_{\rm sky}}}

\def\nfw{(\theta_b)}
\def\fw{\theta_b}
\def\ba{\begin{eqnarray}}
\def\ea{\end{eqnarray}}

\title{Lensing-induced morphology changes in CMB temperature maps in modified gravity theories}

\author[a]{D. Munshi,}
\author[b,c]{B. Hu,}
\author[d]{T. Matsubara,}
\author[a]{P. Coles,}
\author[e]{A. Heavens}

\affiliation[a]{Astronomy Centre, School of Mathematical and Physical Sciences, University of Sussex, Brighton BN1 9QH, U.K.}
\affiliation[b]{Institut de Ci{\`e}ncies del Cosmos (ICCUB), Universitat de Barcelona (IEEC-UB), Mart{\'\i} i Franqu{\`e}s 1, E08028 Barcelona, Spain}
\affiliation[c]{Instituut-Lorentz Theoretical Physics, Universiteit Leiden, Niels Bohrweg 2, 2333 CA}
\affiliation[d]{Kobayashi-Maskawa Institute, Nagoya University, Chikusa, Nagoya 464-8602, JAPAN}
\affiliation[e]{Imperial Centre for Inference and  Cosmology, Blackett Laboratory, Prince  Consort Road, London SW7 2AZ, UK}

\emailAdd{D.Munshi@sussex.ac.uk}
\emailAdd{binhu@icc.ub.edu}
\emailAdd{taka@kmi.nagoya-u.ac.jp}
\emailAdd{P.Coles@sussex.ac.uk}
\emailAdd{a.heavens@imperial.ac.uk}
\abstract
{Lensing of the Cosmic Microwave Background  (CMB) changes the morphology of pattern of temperature fluctuations, so
topological descriptors such as Minkowski Functionals can probe the gravity model responsible for the lensing. 
We show how the recently introduced two-to-two and three-to-one kurt-spectra (and their associated correlation functions),
which depend on the power spectrum of the lensing potential, can 
be used to probe modified gravity theories such as $f({R})$ theories of gravity and quintessence models.
We also investigate models based on effective field theory,  which include the constant-$\Omega$ model, and low-energy Ho\v rava theories.
Estimates of the cumulative signal-to-noise for 
detection of lensing-induced morphology changes, reaches ${\cal O}(10^3)$ for the future planned CMB polarization mission COrE$^{+}$.
Assuming foreground removal is possible to $\ell_{max}=3000$, we show that many modified gravity theories
can be rejected with a high level of significance, making this technique comparable in power to galaxy weak lensing or redshift surveys.
These topological estimators are also useful in distinguishing {\em lensing} 
from other scattering secondaries at the level of the four-point function or trispectrum. Examples include the 
kinetic Sunyaev-Zel'dovich (kSZ) effect which shares, with lensing, a lack of spectral distortion.
We also discuss the complication of foreground contamination from unsubtracted point sources.} 
\begin{document}
\maketitle

\section{Introduction}
\label{sec:Intro}
The all-sky multi-frequency Cosmic Microwave Background (CMB) missions, such as WMAP\footnote {http://map.gsfc.nasa.gov/}, 
Planck\footnote {http://www.rssd.esa.int/index.php?project=Planck}\cite{PC} and
further in future the proposed Experimental Probe of Inflationary Cosmology (EPIC)
survey or ESAs Cosmic Origin Explorer (COrE, \cite{core}), a fourth generation CMB satellite mission concept, are very important in furthering our knowledge of the Universe. 
The
current generation of ground-based observations, namely the Atacama Cosmology 
Telescope (ACT; see ref.\cite{Niem} for ACTPol)\footnote{http://www.physics.princeton.edu/act/} as well as the South Pole Telesecope 
(SPT; see ref.\cite{Jmc} for SPTPol)\footnote {http://pole.uchicago.edu/} are already
providing important clues especially of the CMB secondary anisotropy at smaller angular scales, below a few arc minutes.  Secondary anisotropies, such 
as the thermal Sunyaev-Zel'dovich (tSZ) 
effect and Integrated Sachs-Wolfe (ISW) effect, tell us about the low-redshift Universe and can be a valuable source of cosmological information. In this paper, we focus on another secondary effect, the gravitational lensing of the CMB by the intervening matter. On the one hand, lensing is a source of nuisance for
probing $B$-mode polarization arising from inflationary gravity waves\cite{SH04,Knox}, but on the other hand, lensing of the CMB
allows us to probe the matter distribution at an intermediate redshift ($z\approx 2$),
beyond the typical reach of galaxy lensing surveys.  The study of CMB lensing can tighten constraints on the contents and dynamics of the Universe,
including the dark energy equation of state, neutrino mass
hierarchy \cite{kaknoSo02,LPPP,PZL09,Hu:2012td} and modified theories of gravity \cite{Planck_MG}.  It is the last effect that is the subject of this paper.

Lensing does not change the total power, but it redistributes power preferentially towards smaller angular scale \cite{AntAnt05}, and the effects are most prominent 
below a few arc minutes.  It is challenging to detect since the lensed field has the same spectrum as the unlensed CMB, and detection through the 
angular power spectrum is difficult.
There are other ways to detect the lensing signal, for example through cross-correlation
with external data sets\cite{SmZaDo00,Hirata}, and more recently internally using CMB 
data alone \cite{Smidt10,Su11,Eng12}, and the most recent results from the Planck collaboration include a $40\sigma$ detection of the lensing potential\citep{Planck_phiphi}. The lensing has a quantitatively similar effect on
CMB polarization spectra which may however be significant
at larger angular scales for magnetic or $B$-mode polarization and is of
considerable observational interest \cite{KKS97,SZ97}; a map of $B$-mode polarization
has recently been released \citep{Planck_Bmap}.

In addition to introducing a characteristic
$B$-mode polarization, lensing generates secondary non-Gaussianity (non-Gaussianity) in temperature 
and polarization. While primordial non-Gaussianity can help
to constrain inflation theory\cite{Bartolo04}, similar studies
for secondaries can provide useful clues to structure formation
scenarios. In the absence of any frequency information, information from non-Gaussianity
is helpful in separating out lensing. Early works in this area 
were carried out in real-space \cite{Ber,KC02,KCM02} or
in the harmonic domain using multispectra \cite{VS02,Kogo06}. In the case of an ideal experiment,
with infinite resolution, one-point statistics such as the PDF, lower-order moments 
will not change due to lensing. This too is related to the fact that lensing does not
create power but simply redistributes it. However, experimental beam smoothing,
or any other artificial smoothing, can introduce non-Gaussianity
in even multispectra. For a Gaussian
lensing potential, non-Gaussianity is introduced by lensing alone only the 
trispectrum at lowest order, whereas coupling of lensing with secondary anisotropies such as the ISW and SZ effects
induces a non-zero bispectrum\cite{GS99a,GS99b, Cooray01b,CoorayHu}. 

Minkowski Functionals (MFs) are morphological descriptors that are commonly used in
studying non-Gaussianity in cosmological datasets \cite{Mecke94,SB97}. 
In the CMB, they have already been applied for the analysis of WMAP 3-year data \cite{Hk08}, Boomerang 
\cite{Natoli10}, and more recently to WMAP 7-year data \cite{HikMat12}.  These studies
use a perturbative expansion to express MFs
in terms of the multi-spectra \cite{Mat10}. In general the MFs 
can be expressed as a function
of one-point (generalised) skewness parameters or their higher order analogues.
In recent papers the concept of one-point moments such as skewness and kurtosis, was generalised to
related power-spectra, skew-spectra and kurt-spectra, which carry more information\cite{MuHe10,Mun11,Smidt10b}. 
This extra information is valuable
in separating out individual contributions to the MFs at a given order, as well as to
keep a control on systematics. The aim of this paper is to extend the results 
of recent work \cite{MCH12,modgrav14} where
lensing-induced mode-coupling of the lensing potential and secondaries were
considered, as well as their effect on the morphology of CMB maps to the next
order, i.e. to the level of the trispectrum. In this paper, we apply these statistics to study their ability to constrain modified gravity (MG)
theories and the dark energy (DE) equation of state. For motivation and other cosmological probes 
of MG theories \citep[see e.g. ref.][and references therein]{modgrav15}.

Following ref.\cite{MSC10}, we will generalise the concept of kurt-spectrum and 
show how they can be used to reconstruct the MFs up to the fourth order. 
Kurt-spectra are useful as they can be used to separate lensing from the other
secondaries such as the kinetic Sunyaev-Zel'dovich (kSZ) effect \cite{RS07}.
Similar analysis for frequency-cleaned tSZ maps and weak lensing observations
were recently reported in ref.\cite{MSJC12} and ref.\cite{MWSC12} respectively.
Reconstructing MFs from individual contributions is also important from a different
perspective: the MFs are model-independent statistics and hence care 
must be taken to avoid any serendipitous detection from yet unexplored source
of non-Gaussianity. Throughout, we will use spherical harmonics as basis as lensing of CMB is sensitive
to lensing potential fluctuations at large angular scale $\ell<100$ and high $\ell$ (Limber's) approximation
is not adequate. 

This paper is organized as follows. 
In \textsection\ref{sec:mg} we review modified theories of gravity in the context of Effective Field Theory.
In \textsection\ref{sec:lensing} we discuss the theoretical aspects of lensing-induced secondary non-Gaussianity
in CMB maps. 
In \textsection\ref{sec:MF} we provide details of MFs and related kurt-spectra
for CMB lensing. 
In \textsection\ref{sec:estim} estimators are developed, that can work with realistic mask, noise
and beam. Finally \textsection\ref{sec:disc} is reserved for discussion of our results and
the conclusions are presented in \textsection\ref{sec:conclu}.
In Appendix \ref{sec:appendA} we provide explicit derivations of the two estimators 
as well as their Gaussian counterparts. In Appendix \ref{sec:appendB} we show how the 
one-point kurtosis are recovered from both kurt-spectra. In Appendix \ref{sec:appendC} we
discuss the possibility of constructing sub-optimal estimators for lensing reconstruction. 
\section{Modified Gravity Scenarios in an Effective Field Theory Framework}
\label{sec:mg}
The effective field theory (EFT) approach to dark energy/modified gravity (DE/MG) was recently proposed ~\cite{Gubitosi:2012hu,Bloomfield:2012ff}. An action is built in the Jordan frame and unitary gauge by considering  the operators which are invariant under time-dependent spatial diffeomorphisms. It is able to unify all of the viable single scalar field theories of DE/MG which have a well defined Jordan frame representation, such as $f(R)$ gravity, quintessence, Horndeski models, {\it etc.} (see~ref.\cite{Clifton:2011jh} for a review of the models). In this approach, the additional scalar degree of freedom representing DE/MG is eaten by the metric via a foliation of space-time into space-like hyper-surfaces. Up to the quadratic order, the action reads
\begin{align}
\mathcal{S}_{\rm EFT} = \int d^4x &\sqrt{-g}  \bigg\{ \frac{m_0^2}{2} \left[1+\Omega(\tau)\right] R + \Lambda(\tau) - c(\tau)\,a^2\delta g^{00} \nn  \\
 &  + \frac{M_2^4 (\tau)}{2} \left( a^2\delta g^{00} \right)^2
 - \frac{\bar{M}_1^3 (\tau)}{2} \, a^2\delta g^{00}\,\delta \tensor{K}{^\mu_\mu}  
    - \frac{\bar{M}_2^2 (\tau)}{2} \left( \delta \tensor{K}{^\mu_\mu}\right)^2 \nn \\
   & - \frac{\bar{M}_3^2 (\tau)}{2} \,\delta \tensor{K}{^\mu_\nu}\,\delta \tensor{K}{^\nu_\mu}
      + m_2^2(\tau)\left(g^{\mu\nu}+n^{\mu} n^{\nu}\right)\partial_{\mu}(a^2g^{00})\partial_{\nu}(a^2g^{00}) \nonumber \\
	  & +\frac{\hat{M}^2(\tau)}{2} \, a^2 \delta g^{00}\,\delta \mathcal{R}+	\ldots \bigg\} + S_{m} [g_{\mu \nu}, \chi_m ],\label{actioneft}
\end{align}
where $R$ is the four-dimensional Ricci scalar, $\delta g^{00}$, $\delta \tensor{K}{^\mu_\nu}$, $\delta \tensor{K}{^\mu_\mu}$ and  $\delta \mathcal{R}$ are respectively the perturbations of the upper time-time component of the metric, the extrinsic curvature and its trace and the  three dimensional spatial Ricci scalar. Finally,  $S_m$ is the matter action. 
Since the choice of the unitary gauge breaks time diffeomorphism invariance, each operator in the action can be multiplied by a time-dependent coefficient; in our convention, $\{\Omega,\Lambda,c, M_2^4,\bar{M}_1^3,\bar{M}_2^2,\bar{M}_2^2,\bar{M}_3^2,m_2^2,\hat{M}^2\}$ are unknown functions of the conformal time, $\tau$, and we will refer to them as EFT functions.
We can read that up to the quadratic order, we only have 9 functions. Furthermore, three of them, namely $\{\Omega,c,\Lambda\}$, are the only functions contributing  both to the dynamics of the background and of the perturbations, while the others  play a role only at level of  perturbations. 
Due to the theoretical degeneracies at the kinematic background level, the philosophy of EFT is to 
fix the cosmic background evolution in {\it a priori} manner, then focus on the linear perturbation 
dynamics which are consistent with the given background history. Fixing the time evolution of 
$H(z)$ and $\dot H(z)$ helps us to reduce two of the background 
EFT functions, normally chosen to be $c, \Lambda$, thus reducing the total number of independent 
EFT functions to seven. 

After writing down the generic formula Eq.(\ref{actioneft}), we can see that the only unknown parts of this action are these EFT functions. There are basically two ways to parametrize them, namely covariant mapping parametrization and phenomenological parametrization. The former one is suitable for studying well-known models, which are written in the covariant formalism, while the latter can be used to study the phenomenological models which are inspired by observation. 

In the action Eq.(\ref{actioneft}), the  extra scalar degree of freedom is hidden inside the metric perturbations. However, in order to study the dynamics of linear perturbations and  investigate the stability of a given model, it is more convenient to make it explicit by means of  the St$\ddot{\text{u}}$kelberg technique i.e. performing an infinitesimal coordinate transformation such that $\tau\rightarrow \tau+\pi$, where the new field $\pi$ is the St$\ddot{\text{u}}$kelberg field, which describes the extra propagating degree of freedom.
Varying the action with respect to the $\pi$-field one obtains a dynamical perturbative equation for the extra degree of freedom which allows direct control of the stability of the theory, as discussed at length in ref.~\cite{Hu:2013twa}.

In refs.~\cite{Hu:2013twa,Raveri:2014cka} the EFT framework has been implemented into CAMB/CosmoMC\footnote{\url{http://camb.info}}~\cite{Lewis:1999bs,Lewis:2002ah} creating the EFTCAMB/EFTCosmoMC patches, which are publicly available\footnote{\url{http://wwwhome.lorentz.leidenuniv.nl/~hu/codes/}} (see ref.~\cite{Hu:2014oga} for technical details).  EFTCAMB evolves the full   equations for linear perturbations without relying on any quasi-static approximation. In addition to the standard matter components ({\it i.e.} dark matter, baryon, radiation and massless neutrinos), massive neutrinos have also been included~\cite{Hu:2014sea}.  As mentioned above, EFTCAMB allows the  study of perturbations in a phenomenological way (usually referred to as \emph{pure} EFT mode), investigating the cosmological implications of the different operators in action~Eq.(\ref{actioneft}). It can also be used to study  the exact dynamics for specific models, after the mapping of the given model into the EFT language has been worked out (usually referred to as \emph{mapping} mode). In the latter case one can treat the background via a designer approach, {\it i.e.} fixing the expansion history and reconstructing the specific model in terms of EFT functions; or full mapping approach, {\it i.e.} one can solve the full background and linear perturbation equations of a particular model. Furthermore, the code has a powerful built-in module that investigates whether a chosen model is  viable,  through a set of  general conditions of mathematical and physical stability. In particular,  the physical requirements include the avoidance of ghost and gradient instabilities for both the scalar and the tensor degrees of freedom. The stability requirements are translated into \emph{viability  priors} on the parameter space when using EFTCosmoMC to interface EFTCAMB with cosmological data, and they can sometimes dominate over the constraining power of data~\cite{Raveri:2014cka}.

In this paper, we select a few models both from the {\it pure} EFT mode and also {\it mapping} mode described above. We choose, for the former, the constant-$\Omega$~\cite{Hu:2015rva} models, and for the latter, the designer quintessence~\cite{Hu:2014oga}, $f(R)$ model~\cite{Song:2006ej,Hu:2013twa} and low-energy Ho\v rava gravity~\cite{Horava:2008ih,Frusciante:2015maa}. In the rest part of this section, we will briefly describe these models.  

\subsection{Pure EFT models}
For the {\it pure} EFT models, we select the constant-$\Omega$ model, which consist in taking a constant value for the conformal coupling $\Omega(a)=\Omega_0^{\rm EFT}$ and requiring the expansion history to be exactly that of the $\Lambda$CDM model. This requirement will then fix, through the Friedmann equations, the time dependence of the operators $c$ and $\Lambda$. 
We emphasize here that the constant-$\Omega$ model is not a simple redefinition of the gravitational constant. In fact the requirement of having a $\Lambda$CDM background with a non-vanishing $\Omega$, that would change the expansion history, means that a scalar field is sourced in order to compensate this change. This scalar field will then interact with the other matter fields and modify the behaviour of cosmological perturbations and consequently the CMB power spectra and the growth of structure. For instance, it is easy to show that in the constant-$\Omega$ model, $c(\tau)=\Omega(\rho_m+P_m)/2$, which is vanishing in general relativity, is non-zero.

Another general remark we would like to make on the models that we consider here, is that they display a radically different cosmology, as they correspond to two different behaviours of the perturbation's effective gravitational constant. 
Viable models, in the $f(R)$ case, correspond to an enhancement of the gravitational constant which in turn results in the amplification of the growth of structure that enhances substantially the lensing of the CMB. In the constant-$\Omega$ model, if $\Omega$ is negative, the model will have an enhanced effective gravitational constant with a phenomenology similar to that of $f(R)$ models. Hereafter, we dubbed it as ``${\rm EFT}_1$''. On the other hand, if $\Omega$ is positive the model will be characterized by a smaller effective gravitational constant resulting in a suppression of the growth and consequently a suppression of the CMB lensing. In this paper, we dubbed it as ``${\rm EFT}_2$''. In details, we fix $\Omega_0^{\rm EFT}=-0.1$ and $+0.1$ for ``${\rm EFT}_1$'' and ``${\rm EFT}_2$'', respectively.

\subsection{Mapping models}
For the mapping models, we select three models, namely the designer quintessence, $f(R)$ model and low-energy Ho\v rava gravity. As demonstrated above, one of the advantages of the EFT approach is its ability to unify the languages which describe the linear dynamics of most of the viable single scalar field DE/MG models. 

Via the mapping procedure, we are allowed to design the functional form of the quintessence potential (with canonical kinetic term) to reproduce the input background evolution. Generally speaking, in the minimally-coupled quintessence model, the effect on the growth factor from the quintessence field is sub-dominant compared with its modification to the background expansion. Physically, this is because the Jeans length of the quintessence field is a super-horizon scale, so, there is no significant clustering effect from the scalar degree of freedom. We suggest ref.\cite{Hu:2014oga} for readers who are interested in the details of this model. In the following calculation, we fix $w_0=-0.9$ and $w_a=0.25$ and refer to it as the ``Q'' model.

On the other hand, in the $f(R)$ gravity, the linear perturbation dynamics are more important than its background kinematics. This is because the effective gravitational constant is enhanced by a factor $4/3$ on the scales which are smaller than the Compton wavelength, $B_0\sim 6f_{RR}H^2/(1+f_R)$, of the scalar field. This will magnify the lensing effect significantly, as we will show later. In the following calculation, we fix $B_0=0.1$. 

The last modified gravity models we selected are the low-energy Ho\v rava models. The basic model was first proposed in ref.\cite{Horava:2008ih} to solve the UV complete problem of quantum gravity, then it was embedded in the EFT approach in ref.\cite{Frusciante:2015maa} to study cosmic late-time acceleration. Basically, its phenomenology on the background is simply rescaling the Hubble parameter; and on the perturbation level, due to the strong coupling with the gravity sector, the scalar field perturbations could suppress the linear structure formation rate substantially. In this paper, we select two Ho\v rava models, one with three parameters ($\lambda = 1.4$, $\xi = 0.9$, $\eta = 1.0$), the other with two ($\lambda = 1.4$, $\eta = 1.0$). Hereafter, we dubbed them as ``${\rm H}_3$'' and ``${\rm H}_2$'' models, respectively. Compared with the  ``${\rm H}_3$'' model, the ``${\rm H}_2$'' are designed to evade the PPN constraint. 

Finally, we take the $\Lambda$CDM model (``$\Lambda$'') as the baseline model, and all the vanilla cosmological parameters are the same as the ones 
from the Planck-2015 \cite{Ade:2015xua} data release.  
 
\section{Lensing induced non-Gaussianity in CMB Temperature Maps}
\label{sec:lensing}
In this section we will briefly review certain aspects of lensing of the CMB \cite{Hu01,huOka02,Kogo06}. For a full review see ref.\cite{AntAnt05}. 
\subsection{Lensing in Temperature Maps}
In the context of CMB lensing, the surface of last scattering
can be thought of as a single source plane. The projected lensing potential $\phi(\oh)$
towards an angular direction $\oh=(\theta,\phi)$ can be expressed in terms of a line of
sight integration of the 3D potential $\Phi(\br)=\Phi(r,\oh)$:
\be
\phi(\oh) = -2\int_0^{r_0} dr {{d_A(r-r)} \over d_A(r) d_A(r_0)}\Phi(r,\oh).
\ee
Here $r_0$ is the comoving conformal distance to the surface of last scattering and $d_A(r)$
is the comoving angular diameter distance out to $r$. Lensing effectively redistributes
the temperature $\Theta(\oh)=[\delta T(\oh)/T_0]$ on the surface of the sky 
through the angular deflections resulting along the photon path,
 $\alpha(\oh)=\nabla\phi(\oh)$ such that $\Theta(\oh) = \bar\Theta(\oh+\alpha)$, where $\Theta$ is the lensed CMB sky and $\bar\Theta$
corresponds to the temperature distribution in the absence of lensing, and $\nabla$ is 
the covariant derivative on the surface of the unit sphere. 
For the following discussion we will ignore the secondary contribution.
Coupling of lensing with secondaries and resulting impact on morphological
properties of CMB maps have been discussed in detail in \cite{MCH12}. 
Expanding the above expression in a Taylor series we can write \cite{GS99a,GS99b}: 
\ben
&& \Theta(\oh) = \bar\Theta(\oh+\alpha) \approx \bar\Theta(\oh)+
\nabla_i\phi(\oh)\nabla^{i}\bar\Theta(\oh)+{\cal O}(\phi^2); \label{eq:expansion}\\
&& \delta \Theta(\oh) = \Theta(\oh) - \bar\Theta(\oh) \approx \nabla_i\phi(\oh)\nabla^{i}\bar\Theta(\oh).
\een
In the harmonic domain, using spherical harmonics $Y_{lm}(\oh)$ as the basis function,
we can express the multipole of lensing induced temperature anisotropy $\delta \Theta_{lm}$
in terms of multipoles of lensing potential $\phi_{LM}$ and multipole of the unlensed 
CMB temperature anisotropy $\bar\Theta_{lm}$ respectively: 
\ben
&& \delta \Theta_{\ell m} = \int d\oh  Y^*_{\ell m}(\oh)\nabla_i\phi(\oh)\nabla^{i}\bar\Theta(\oh) \nn \\
&&  = \int d\oh Y^*_{\ell m}(\oh)\nabla_i 
\left [\sum_{LM}\phi_{LM}Y^*_{LM}(\oh) \right ] 
\nabla^{i}\left [\sum_{\ell'm'}\bar\Theta_{\ell'm'}Y^*_{\ell'm'}(\oh) \right ]\nn \\
&& = \sum_{LM}\sum_{\ell'm'} (-1)^m \phi_{LM}\bar\Theta_{\ell'm'} \; {}_0F_{\ell\ell'L}
\left ( \begin{array}{ c c c }
     \ell & \ell' & L \\
     m & -m' & -M
  \end{array} \right); 
\label{eq:harmonics_temp}
\een
where we have used the Gaunt integral \cite{Ed68} to arrive at the last line. The following
notations were introduced:
\ben
&& {}_{\pm s}F_{\ell L \ell'} \equiv {1 \over 2} {}_{\pm s}I_{\ell\ell'L} \Lambda_{\ell\ell'L} = 
{1 \over 2}{}_{\pm s}I_{\ell\ell'L}[\Pi_{L}+\Pi_{\ell'}-\Pi_{\ell}]; \quad\\
&& {}_{\pm s}I_{\ell\ell'L} = \sqrt{\Pi_{\ell}\Pi_{\ell'}\Pi_{L} \over 4\pi}
\left ( \begin{array}{ c c c }
     \ell & \ell' & L \\
     \pm s & \mp s &  0
  \end{array} \right),
\label{eq:f_pm2}
\een
\n
where $\Pi_\ell\equiv \ell(\ell+1)$ and the last matrix is a Wigner $3j$ symbol. The function ${}_{\pm s}F_{l_1Ll_2}$ encodes the rotationally-invariant
part of the coupling between three multipoles $(l_1,l_2,L)$. To simplify our notation, as is common in
the literature, we will denote ${}_{0}I_{\ell\ell'\ell''}$ as
$I_{\ell\ell'\ell''}$ by dropping the $s$ index.

The analytical modelling of higher-order correlation functions 
is most naturally done in the harmonic domain where they
are represented by the multi-spectra. It is known that
lensing of CMB only induces even-order multi-spectra. 
Thus the lowest-order departure
from Gaussianity, in case of CMB lensing, is characterized
by the connected part of the four-point correlation
function (or equivalently the trispectrum in the harmonic domain) defined through the relation:
\ben
\la\Theta_{\ell_1m_1}\Theta_{\ell_2m_2}\Theta_{\ell_3m_3}\Theta_{\ell_4m_4} \ra =
\la\Theta_{\ell_1m_1}\Theta_{\ell_2m_2}\Theta_{\ell_3m_3}\Theta_{\ell_4m_4} \ra_{\rm c} +
\la\Theta_{\ell_1m_1}\Theta_{\ell_2m_2}\Theta_{\ell_3m_3}\Theta_{\ell_4m_4} \ra_{\rm G};
\label{eq:4pt}
\een
where the subscripts $_G$ and $_c$ correspond to Gaussian and non-Gaussian
(or connected) contributions to the four-point correlation function. The connected
part of the four-point correlation function in real-space is related to
the trispectrum $ T^{\ell_1\ell_2}_{\ell_3\ell_4}(\ell)$ through the following relation:
\ben
&& \la\Theta_{\ell_1m_1}\Theta_{\ell_2m_2}\Theta_{\ell_3m_3}\Theta_{\ell_4m_4} \ra_c \nn = \\
&& \sum_{\ell m} (-1)^m T^{\ell_1\ell_2}_{\ell_3\ell_4}(\ell)\left ( \begin{array}{ c c c }
     \ell_1 & \ell_2 & \ell \\
     m_1 & -m_2 & -m
  \end{array} \right)
\left ( \begin{array}{ c c c }
     \ell_3 & \ell_4 & \ell \\
     m_3 & -m_4 & -m
  \end{array} \right).
\label{eq:def_tri}
\een
To impose the symmetry inherent in the trispectrum it is expressed in terms of its ``pairing matrix'' $P^{\ell_1\ell_2}_{\ell_3\ell_4}(\ell)$.
\ben
T^{\ell_1\ell_2}_{\ell_3\ell_4}(\ell) =  P^{\ell_1\ell_2}_{\ell_3\ell_4}(\ell) 
&& + \Xi_{\ell}\Big [\sum_{l'} (-1)^{\ell_2+\ell_3}\left \{ \begin{array}{ c c c }
     \ell_1 & \ell_2 & \ell \\
     \ell_4 & \ell_3 & \ell'
  \end{array} \right \} P^{\ell_1\ell_3}_{\ell_2\ell_4}(\ell') \nn \\
&& + \sum_{\ell'}  
(-1)^{L+L'} \left \{ \begin{array}{ c c c }
     \ell_1 & \ell_2 & \ell \\
     \ell_3 & \ell_4 & \ell'
  \end{array} \right \} P^{\ell_1\ell_4}_{\ell_3\ell_2}(\ell') \Big ],
\label{Total_Tri}
\een
where $ \Xi_{\ell} \equiv 2\ell+1$, and the matrices in curly brackets are Wigner $6j$-symbols which are defined in terms of $3j$ symbols 
(see ref.\cite{Ed68}). 
The ``pairing matrix''  can be further decomposed in terms of the
{\it reduced} trispectrum $\tau^{\ell_1\ell_2}_{\ell_3\ell_4}(\ell)$:
\ben
&& P^{\ell_1\ell_2}_{\ell_3\ell_4}(\ell) = \tau^{\ell_1\ell_2}_{\ell_3\ell_4}(\ell)
+ (-1)^{\Sigma_U}\tau^{\ell_2\ell_1}_{\ell_3\ell_4}(\ell) + (-1)^{\Sigma_L} \tau^{\ell_1\ell_2}_{\ell_4\ell_3}(\ell) + 
(-1)^{\Sigma_L + \Sigma_U}\tau^{\ell_2\ell_1}_{\ell_4\ell_3}(\ell);\nn\\
&& \quad \Sigma_L = \ell_1+\ell_2+\ell; \quad \Sigma_U = \ell_3+\ell_4+\ell.
\een
In the case of weak lensing of CMB the reduced trispectrum $\tau$ depends only on the
power-spectrum of the lensing potential $\myC_{\ell}^{\phi\phi}=\la\phi_{\ell m}\phi^*_{\ell m}\ra$ and the power spectrum of temperature anisotropy $\myC^T_{\ell}$:
\be
\tau^{\ell_1\ell_2}_{\ell_3\ell_4}(\ell)=\myC_{\ell}^{\phi\phi}\bar\myC^{T}_{\ell_2}\bar\myC^{T}_{\ell_4} \; 
{}_0F_{\ell_1\ell\ell_2} \;{}_0F_{\ell_3\ell\ell_4}
\label{eq:spin0_tau}
\ee
Previous studies have already shown that the pairing matrix $P^{\ell_1\ell_2}_{\ell_3\ell_4}(\ell)$
can very accurately describe the trispectrum $\tau^{\ell_1\ell_2}_{\ell_3\ell_4}(\ell)$:
\ben
T^{\ell_3\ell_4}_{\ell_1\ell_2}(\ell) \approx P^{\ell_3\ell_4}_{\ell_1\ell_2}(\ell) &=& \myC_{\ell}^{\phi\phi}
\left ( \bar\myC^{T}_{\ell_2} {}_0F_{\ell_1\ell\ell_2} + \bar\myC^{T}_{\ell_1} {}_0F_{\ell_2\ell\ell_1} \right )
\left ( \bar\myC^{T}_{\ell_4} {}_0F_{\ell_3\ell\ell_4} + \bar\myC^{T}_{\ell_3} {}_0F_{\ell_4\ell\ell_3} \right )\\
&=& \myC_\ell^{\phi\phi} f_{\ell_1\ell\ell_2}f_{\ell_3\ell\ell_4}
\een
This is the approximation which we will use in our study.
\begin{figure}
\vspace{1.25cm}
\begin{center}
{\epsfxsize=13. cm \epsfysize=5. cm 
{\epsfbox[32 411 547 585]{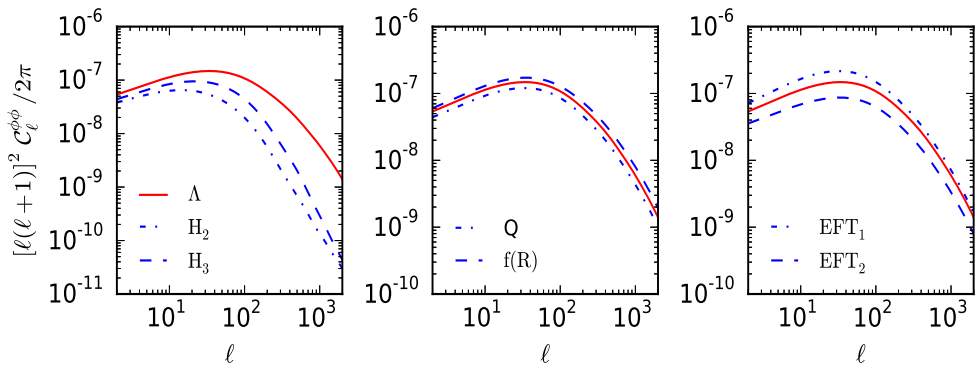}}}
\caption{CMB Lensing potential power sepctrum $\myC^{\phi\phi}_{\ell}$ as a function of $\ell$ for various Modified Gravity theories.
In left panel we show two Ho\v rava theories denoted as $\rm H_2$ (dot-dashed line) and $\rm H_3$ (dashed line). 
The standard $\Lambda$CDM prediction is depicted as a solid-line in each panel.
In the middle panel we compare results for a quintessence model (dot-dashed) and $f(R)$ theory (dashed) results. In the right panel we show 
results from two different EFT caclulations EFT$_1$ (dot-dashed) and EFT$_2$ (dashed).  See text for more details.}
\label{fig:phiphi}
\end{center}
\end{figure}
\subsection{Gaussian Component} The Gaussian 
component of the four-point correlation function defined in Eq.(\ref{eq:4pt}) can also be expressed
as follows:
\ben
&&\la\Theta_{\ell_1m_1}\Theta_{\ell_2m_2}\Theta_{\ell_3m_3}\Theta_{\ell_4m_4} \ra_{\rm G} \nn \\
&& = \sum_{\ell m} (-1)^m G_{\ell_3\ell_4}^{\ell_1\ell_2}(\ell)
\left ( \begin{array}{ c c c }
     \ell_1 & \ell_2 & \ell \\
     m_1 & -m_2 & -m
  \end{array} \right )
\left ( \begin{array}{ c c c }
     \ell_3 & \ell_4 & \ell \\
     m_3 & -m_4 & -m
  \end{array} \right )
\label{eq:gauss_tri}
\een

The Gaussian component of the trispectrum defined above $G^{{\ell}_1{\ell}_2}_{{\ell}_3{\ell}_4}({\ell})$ representing
the disjoint contribution to our-point correlation function is determined completely by the
(unlensed) CMB power spectrum $\bar \myC_{\ell}^{T}$:
\ben
G^{\ell_1\ell_2}_{\ell_3\ell_4}(\ell)&=&
(-1)^{\ell_2+\ell_3} \sqrt {\Xi_{\ell_1}\Xi_{\ell_3}} \myC^T_{\ell_1}\myC^T_{\ell_3}
\delta_{\ell_1\ell_3}\delta_{\ell_2\ell_4}\delta_{\ell 0} \nn\\
&& +\Xi_{\ell}\;\myC^T_{\ell_1}\myC^T_{\ell_2}\left [ (-1)^{\ell_1+\ell_2+\ell}\delta_{\ell_1\ell_3}\delta_{\ell_2\ell_4} 
+\delta_{\ell_1\ell_4}\delta_{\ell_2\ell_3} \right ] 
\label{eq:def_gauss}
\een
Henceforth, we will ignore the $\ell=0$ mode as it does not contribution to the deflection of photons.

\subsection{Foregrounds} In addition to cosmological sources of non-Gaussianity both primary and secondary,
foreground such as the unsubtracted point-sources can also make a significant contribution.

The bispectrum and trispectrum from extragalactic radio and infra-red sources with fluxes ${F}$ smaller than a certain
detection threshold ${F}_{\rm d}$ is simple to estimate if Poisson distributed. The trispectrum for the unsubtracted point source distribution then has a constant amplitude $t_{\rm ps}$:
\be
T^{\ell_3\ell_4}_{\ell_1\ell_2}(\ell) =t_{\rm ps}\, I_{\ell_1\ell_2\ell}I_{\ell_3\ell_4\ell}.
\label{eq:unresolved}
\ee
\begin{figure}
\vspace{1.25cm}
\begin{center}
{\epsfxsize=13. cm \epsfysize=5. cm 
{\epsfbox[32 411 547 585]{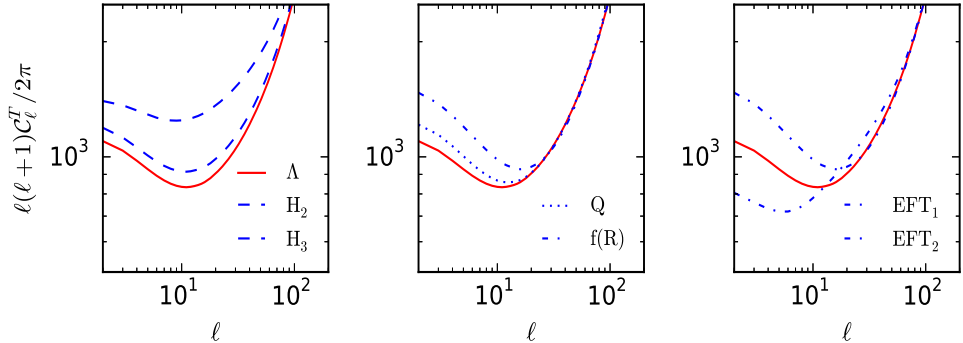}}}
\caption{CMB temperature power spectra for the models shown in Figure-\ref{fig:phiphi}. The line styles
representing various models remain the same.}
\label{fig:isw}
\end{center}
\end{figure}
Following the procedure outlines in ref.\cite{Bartolo04} we obtain the following expression for $t^{\rm ps}$:
\ben
t_{\rm ps}= {(2-\beta)^2 \over \beta (4-\beta)} [n(<{F}_{\rm d})]^{-1} [{\cal C}_{\rm ps}]^2.
\een
Here ${\cal C}_{\rm ps}$ is the $\ell$ independent power spectrum for the point sources which has the following expression:
\ben
{\cal C}_{\rm ps}= g^2(x) {\beta \over 2-\beta} n(<{F}_{\rm d}) {F}_{\rm d}^2.
\een
Here $d{\rm n}/d{\rm F}$ is the differential source count per unit solid angle and we have defined 
$n(<F_d) = \int_0^{F_d} d{F} {d{\rm n}/d{F}}$.
It is generally assumed to be a power-law, $dn/{dF} \propto {F}^{-\beta-1}$.
For example, for Euclidean source counts $\beta=3/2$. We have defined $x={h\nu/k_B {\rm T}} \approx (\nu/56.80 {\rm GHz})({\rm T}/2.726)^{-1}$
and $g(x)= 2\,(hc)^2/({k_B \rm T})^3[{\rm sinh} (x/2)/ x^2]^2$.
For the 217 GHz assuming $n(<{F}_{\rm d})=100$ we obtain $t_{\rm ps} \approx 2\times 10^{-38}$ 
and for $90$ GHz assuming $n(<{F}_d)=7$ 
we get $t_{\rm ps} \approx 2\times 10^{-34}$. These results should only be considered a very crude order of magnitude
estimates.
\section{Morphological Estimators}
\label{sec:MF}
We outline our estimators in this section and relate them to morphological statistics such as the Minkowski Functionals. 
\subsection{Minkowski Functionals}
The MFs are well known morphological descriptors which are used in the study of random fields. 
Morphological properties are the properties that remain invariant under rotation and translation (see ref.\cite{Hadwiger59}
for more formal introduction). They are defined
over an excursion set $\Sigma$ for a given threshold $\nu$. The three MFs that 
we will use for two dimensional (2D) temperature anisotropy
$\Theta(\oh)$ defined on the surface of the sky can be expressed as:
\be
{\cal V}_0(\nu) = \int_{\Sigma} da; \quad {\cal V}_1(\nu) = {1 \over 4}\int_{\partial\Sigma} dl; \quad {\cal V}_2(\nu) = {1 \over 2\pi}\int_{\partial \Sigma}\kappa dl; \quad \nu = {\Theta \over \sigma_0}. 
\ee
\n
Here $da$, $dl$ are the elements for the excursion set $\Sigma$ and its boundary $\partial \Sigma$. The MFs ${\cal V}_k(\nu)$
correspond to the area of the excursion set $\Sigma$, the length of its boundary $\partial\Sigma$ as well as the
integral curvature along its boundary which is related to the genus $g$ and hence the Euler characteristics $\chi$.

The MFs for a random Gaussian field are well known and given by Tomita's formula \cite{Tom86} and are completely defined by the corresponding power spectrum 
$\myC_\ell$. A perturbative analysis was suggested to go beyond the Gaussian distribution 
in ref.\cite{Mat10}:
\ben
&& {\cal V}_k(\nu) = 
{1 \over (2\pi)^{(k+1)/2}} {\omega_2 \over \omega_{2-k}\omega_k} \exp \left ( -{\nu^2 \over 2}\right ) \left ( \sigma_1 \over \sqrt 2 \sigma_0 \right )^k v_k(\nu); \\
&& v_k(\nu) = \left [ v_k^{(0)}(\nu)+v_k^{(1)}(\nu)\sigma_0 + v_k^{(2)}(\nu)\sigma_0^2 + \cdots \right ]; \\
&& \sigma_j^2 = {1 \over 4\pi }\sum_\ell \,\Xi_{\ell}\, \Pi_{\ell}^j\, \myC_\ell b_\ell^2(\theta_0);\\
&& {\Pi}_{\ell} = \ell(\ell+1); \quad b_\ell(\theta_0) = \exp\left [-\Pi_{\ell}{\theta_b^2} \right ];
\quad \theta_b = {\theta_0\over \sqrt{16\ln 2}}.
\label{eq:v_k}
\een
\ben
&& H_{-1}(\nu) = \sqrt{\pi \over 2} e^{\nu^2/2} {\rm erfc} \left({\nu \over 2} \right); \\
&& H_{0}(\nu) = 1; \quad H_1(\nu) = \nu; \quad H_2(\nu)=\nu^2-1;  \\
&& H_3(\nu)= \nu^3-3\nu; \quad H_4(\nu) = \nu^4 - 6\nu^2 +3; \\
&& H_n(\nu) = (-1)^n \exp \left ( {\nu^2 \over 2} \right) {d^n \over d\nu^n} \exp \left(-{\nu^2 \over 2} \right ).
\een

\n
We have assumed a Gaussian beam $b_\ell(\theta_0)$ with FWHM $=\theta_0$. The constant $\omega_k$ introduced above is the volume of the unit sphere in k-dimension. $w_k = {\pi^{k/2}/ \Gamma(k/2+1)}$
in 2D we will only need $\omega_0=1$, $\omega_1=2$ and $\omega_2 =\pi$. For a 
purely Gaussian distribution  $v_k^{(0)}(\nu)=H_{k-1}(\nu)$ and all higher order terms
vanish. We notice that $\sigma_0^2 = \la\delta\Theta^2\ra$ and $\sigma_1^2 = \la|\nabla\Theta|^2\ra$.
\subsection{Kurtosis Spectra}
The leading order terms that signify non-Gaussianity 
of MFs depend on the bispectrum or equivalently a set of three generalised skewness parameters. 
The next to the leading order order correction terms depend on a set of four {\em generalised} 
kurtosis parameters $K^{(i)}$ that are fourth order statistics. In general the kurtosis parameters are collapsed fourth order
one-point cumulants and probe the trispectrum with varying weights \cite{Mu_kurt10}. 
The four different kurtosis parameters 
that are related to the MFs are a natural generalisation of the ordinary kurtosis $K^{(0)}$ which
is routinely applied in many cosmological studies. We will denote these generalised kurtosis parameters by $K^{(i)}; i = 1,2,3$.
These parameters are constructed from the derivative field of the original map
$\Theta(\oh)$ and its derivatives  $|\nabla_{i}\Theta(\oh)|^2 = [\nabla_{i}\Theta(\oh)\nabla^{i}\Theta(\oh)]$ and $[\nabla^2\Theta(\oh)]$.
\ben
&& K^{(0)} \equiv  {\la \myf^4 \ra_c \over \sigma_0^6}; \quad
K^{(1)} \equiv   {\la \myf^3 \nabla^2 \myf \ra_c \over \sigma_0^4 \sigma_1^2}; \quad \\
&& K^{(2)} \equiv  K^{(2a)} +   K^{(2b)} \equiv 
 2 {\la \myf |(\nabla \myf)|^2 (\nabla^2\myf) \ra_c  \over \sigma_0^2\sigma_1^4}
+ {}{\la |(\nabla \myf)|^4 \ra_c  \over \sigma_0^2 \sigma_1^4};\quad
 K^{(3)} \equiv {\la |\nabla\myf|^4 \ra_c \over 2\sigma_0^2\sigma_1^4}.
\label{kurtosis_real_space}
\een
The subscript $_c$ correspond to the connected components which indicates that all Gaussian unconnected contributions
are subtracted out, these include both noise as well as the signal contribution. 

If we ignore lensing-secondary coupling contributions discussed in ref.\cite{MCH12} and contribution from primordial
non-Gaussianity\cite{Mu_kurt10}, the next-to-leading order corrections to the MFs involve 
tri-spectral contributions $K^{(i)}$s which can be derived following ref.\cite{Mat10}. 
\ben
&& v_0^{(4)}(\nu) = {K^{(0)} \over 24} {\cal H}_3(\nu); \quad \label{one_pt0} \\
&&  v_1^{(4)}(\nu) ={K^{(0)} \over 24} {\cal H}_4(\nu) -{ K^{(1)}\over 12}  {\cal H}_2(\nu)
- { K^{(3)} \over 8} \quad \label{one_pt1}\\
&& v_2^{(4)}(\nu) =  {K^{(0)} \over 24} {\cal H}_5(\nu) 
 -{  K^{(1)} \over 6}{\cal H}_3(\nu) - {  K^{(2)} \over 2} {\cal H}_1(\nu). \label{one_pt2}
\label{eq:fourth_mink}
\een

\n  
Next, we will introduce three additional trispectra that are constructed using different weights to the original beam-smoothed trispectra 
$T$ and differ in the way they weight various modes, which are specified by a particular choice of the quadruplet of angular harmonics $\{ \ell_i \}$:
\ben
&& T^{(0)} =  {T \over \sigma_0^6}; \label{t0}\quad\\
&& T^{(1)} = -{1 \over 4 \sigma_0^4 \sigma_1^2} 
\left [{{\Pi_{\ell_1}}+{\Pi}_{\ell_2}+{\Pi}_{\ell_3}+{\Pi}_{\ell_4}}\right ] T; \\
&& T^{(2)} = {1 \over 4\sigma_0^2 \sigma_1^4}\Big[{ {\Pi}^2_{L}- ({\Pi}_{\ell_1}+{\Pi}_{\ell_2})}
{({\Pi}_{\ell_3}+{\Pi}_{\ell_4}) }\Big ] T; \\
&& T^{(3)} = {1 \over 4 \sigma_0^2\sigma_1^4}\left [ {({\Pi}_{\ell_1}+ {\Pi}_{\ell_2}-{\Pi}_{L})
({\Pi}_{\ell_3}+{\Pi}_{\ell_4}-{\Pi}_{L})}\right  ] T; \label{t4}\\
&& T \equiv T^{\ell_1\ell_2}_{\ell_3\ell_4}(L)b_{\ell_1}(\theta_0)b_{\ell_2}(\theta_0)
b_{\ell_3}(\theta_0)b_{\ell_4}(\theta_0) 
\label{eq:kurt_spectra1}
\een
We can define similar expressions for the Gaussian component simply by replacing the connected part of the
trispectrum $T\equiv T^{\ell_1\ell_2}_{\ell_3\ell_4}(\ell)$ (defined in Eq.(\ref{eq:def_tri})) with the disconnected part of the trispectrum
$G\equiv  G^{\ell_1\ell_2}_{\ell_3\ell_4}(\ell)$ 
(defined in Eq.(\ref{eq:def_gauss}))
which will be useful for constructing the disconnected part of the kurt-spectra that
we need to subtract to retain only the non-Gaussianity part.

\begin{figure}
\vspace{1.25cm}
\begin{center}
{\epsfxsize=16. cm \epsfysize=5.5 cm 
{\epsfbox[75 451 567 602]{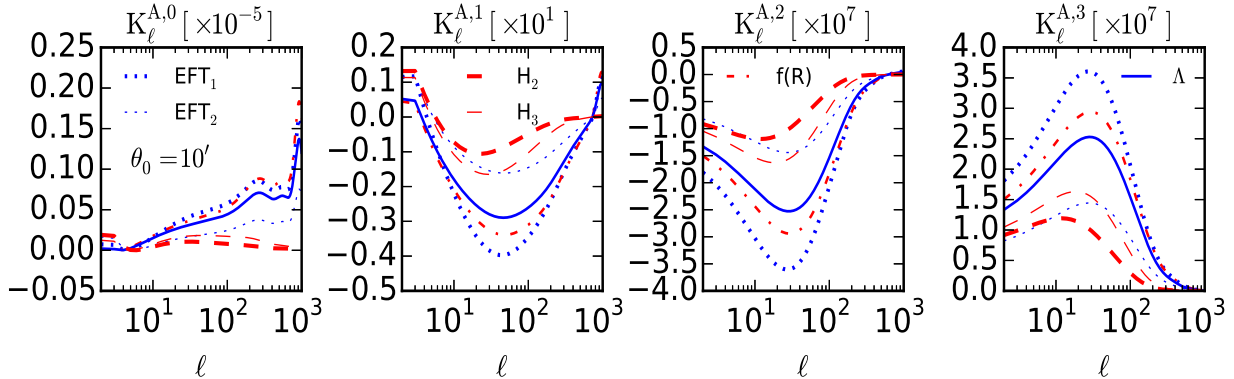}}}
\caption{The kurtosis spectra $K^{A,i}_{\ell}$ defined in Eq.(\ref{eq:kurt_spectra2a}) for smoothing
angular scale $\theta_0=10^{\prime}$ and $\ell_{max}=10^3$. From left they correspond
to $K^{A,1}_{\ell}$ through to  $K^{A,4}_{\ell}$. In each panel the base GR+$\Lambda$CDM
model is shown along with five other modified gravity theories. The two effective field theory
models are depicted by dashed line. The thick dotted lines correspond to the EFT$_1$ and the thin
dotted lines correspond to EFT$_2$ respectively. The two dashed lines correspond to the Ho\v rava models.
The thick (thin) solid lines correspond to H$_2$(H$_3$). The doted lines correspond to the $f(R)$ model
(see text for the description of these models). We haven't included the $\sigma_0(\theta_0)$ and $\sigma_1(\theta_0)$ 
dependent normalisations in these plots to isolate the effect of MG theories on trispectrum.}
\label{fig:A31}
\end{center}
\end{figure} 

These results are based on the following properties of the spherical harmonics:
\ben
&& \int \nabla_i Y_{\ell_1m_1}(\oh) \nabla^{i} Y_{\ell_2m_2}(\oh) Y_{LM}(\oh)d\oh = 
{1 \over 2}({\Pi}_{\ell_1} + {\Pi}_{\ell_2} -{\Pi}_{L}) I_{\ell_1\ell_2L}
\left ( \begin{array}{ c c c }
     \ell_1 & \ell_2 & L \\
     m_1 & m_2 & M
  \end{array} \right )\\
&& \nabla^2 Y_{\ell m}(\oh) = -\Pi_{\ell}  Y_{\ell m}(\oh)
\een 
\begin{table}
\begin{center}
\caption{Survey Parameters \cite{mission}}
\vspace{0.5cm}
\begin{tabular}{| c | c | c | c | c |c }
  \hline
  Mission &  Frequency (GHz) & Sensitivity ($\mu$K-arcmin) & $f_{\rm sky}$  & FWHM \\
  \hline
  \rowcolor[gray]{0.8}  COrE$^+$ & 145 & $5.0$ & 70\% & 5.8$^{\prime}$ \\
 \hline
ACTPol & 150 & $9.8$ & 50\%& 1.3$^{\prime}$\\
\hline
\rowcolor[gray]{0.8}  Planck  & 143 & $44.0$& 50\%& 7.3$^{\prime}$\\
\hline
\end{tabular}
\label{tabular:tab1}
\end{center}
\end{table}

Following the prescriptions in ref.\cite{Mu_kurt10} and ref.\cite{MCCHS11}, the four generalised kurtosis $K^{(i)}$, which are one-point
statistics, the concept of {\em two-to-two} $K^{A,i}_\ell$ and {\em three-to-one}
$K^{B,i}_\ell$ kurt-spectra can now be introduced in terms of the
generalised tri-spectra $T^{(i)}$ as follows:
\ben
&& K^{A,i}_\ell = \sum_{\ell_i} [T^{(i)}]^{\ell_1\ell_2}_{\ell_3\ell_4}(\ell) J_{\ell_1\ell_2\ell} J_{\ell_3\ell_4\ell};
\label{eq:kurt_spectra2a}\\
&& K^{B,i}_\ell = \sum_{\ell_i}\sum_{L} [T^{(i)}]^{\ell_1\ell_2}_{\ell_3\ell}(L) J_{\ell_1\ell_2L} J_{L\ell_3\ell}; 
\label{eq:kurt_spectra2b}\\
&& J_{\ell_1\ell_2\ell_3} = {1 \over \Xi_{\ell_3}}I_{\ell_1\ell_2\ell_3}.
\een
These estimators generalises the {\em optimized} version of $K^{A,0}_\ell$ that has already been used in ref.\cite{Smidt10} to constrain
the projected mass exploiting the fact that the estimators $K^{A,0}_\ell$ and its higher order
analogues are directly proportional to the lensing power-spectrum $\myC_\ell^{\phi\phi}$.

The main advantage of using two-point estimators such as $K^{A,i}_\ell$ and $K^{B,i}_\ell$
is in the additional information contained in the shape of these
spectra, which is useful in differentiating them from other secondary contributions
e.g. kSZ.
\begin{figure}
\vspace{1.25cm}
\begin{center}
{\epsfxsize=16. cm \epsfysize=5.5 cm 
{\epsfbox[75 451 567 602]{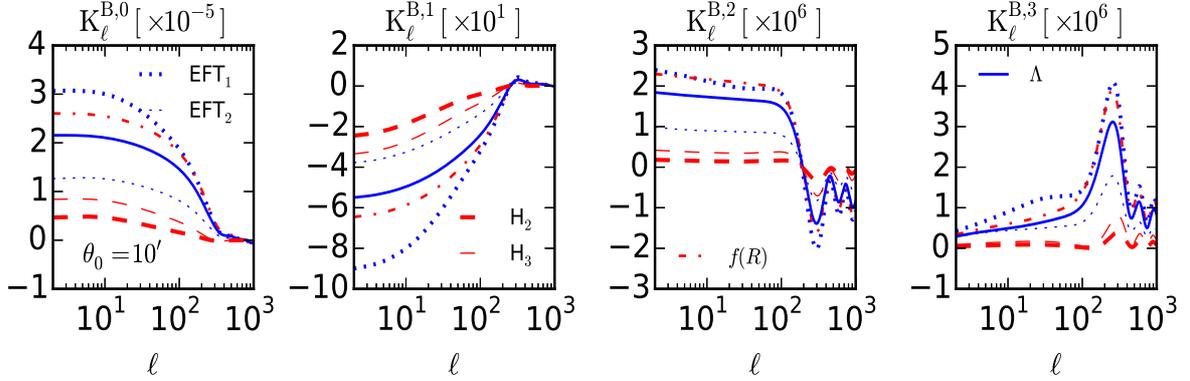}}}
\caption{Same as previous figure but for the power-spectra $K^{B,i}_{\ell}$ as defined in Eq.(\ref{eq:kurt_spectra2b}).}
\label{fig:B22}
\end{center}
\end{figure}
The one-point statistics which are used as an input in Eq.(\ref{one_pt1}) and Eq.(\ref{one_pt2})
to construct the MFs, can be computed from their two-point counterparts:
\be
K^{(i)} \equiv {1\over 4\pi}\sum_{\ell} \Xi_{\ell}\, K^{A,i}_\ell ={1 \over 4\pi}\sum_{\ell} \Xi_{\ell}\, K^{B,i}_\ell.
\label{eq:one_pt}
\ee 
The physical meaning of these kurt-spectra can be understood more easily 
in the harmonic domain. Each individual mode of the trispectrum
is characterized by a specific choice of set of modes ${\ell_i}$ that defines it. 
These modes each constitute the sides of a quadrilateral  
whose diagonal is specified by the harmonics $\ell$. The kurt-spectra $K_\ell^{A,i}$ 
considered here take contributions from all possible configurations of the 
quadrangle representing trispectrum 
while keeping its diagonal $\ell$ fixed. The kurt-spectra  $K_\ell^{B,i}$
on the other hand represent the sum over all possible configurations of the 
quadrangle while keeping one of its side $\ell$ fixed.

The estimation of the kurt-spectra from real data is relatively easy and follows the same
methodology as that of the skew-spectra. The first of these kurt-spectra $K^{(0)}$ is extracted by cross-correlating
the squared field $[\myf^2(\oh)]$ with itself. The spectra  $K^{(1)}$ is constructed by
cross-correlating $[\myf^2(\oh)]$ against $[\myf(\oh)\nabla^2\myf(\oh)]$. The other two kurt-spectra can
likewise be constructed. In each such construction a scalar map from a product field is 
generated before it is cross-correlated with another such map. 
\n
The explicit expressions for the estimators $ \hat K^{A,i}_\ell$ are as follows:  
\ben
&& \hat K^{A,0}_\ell =  
A_0\,\hat K^{\Theta^2,\Theta^2}_\ell; \quad
  \hat K^{A,1}_\ell = 
A_1\, \hat K^{\Theta^2,\Theta\nabla^2\Theta}_\ell\label{eq:k0}\\
&&  \hat K^{A,2}_\ell = 
A_2\, (2 K_\ell^{\Theta\nabla^2\Theta,\nabla\Theta\cdot\nabla\Theta} 
+ K_\ell^{\nabla\Theta\cdot\nabla\Theta,\nabla\Theta\cdot\nabla\Theta});
\quad \hat K^{A,3}_\ell = A_3\, K^{\nabla\Theta\cdot\nabla\Theta,
\nabla\Theta\cdot\nabla\Theta}_\ell; \\
&& A_0= {\sigma_0^{-6}}; \quad A_1 = {\sigma_0^{-4}\sigma_1^{-2}};\quad  A_2 =A_3 = {\sigma_0^{-2}\sigma_1^{-4}}.
\label{eq:k4}
\een

For current-generation surveys with small sky coverage, the correlation functions associated with the kurt-spectra may have some practical advantages.  These are
\ben
&&  K^{A,i}_{12}[\theta] \equiv {1 \over 4\pi}\sum_{\ell}\Xi_{\ell} P_{\ell}(\cos\theta)\,K^{A,i}_{\ell}\; \label{eq:corrA}\\
&& K^{B,i}_{12}[\theta] \equiv {1\over 4\pi}\sum_{\ell}\Xi_{\ell} P_{\ell}(\cos\theta)\,K^{B,i}_{\ell}. \label{eq:corrB}
\een
Here $P_{\ell}$ is a Legendre polynomial of order $\ell$. These two-point correlation functions
(also known as cumulant correlators) are defined on the surface of the sphere between two line-of-sight directions separated 
by an angle $\theta$. For the special case of $\theta=0$ they both collapse to the same one-point
cumulants defined in Eq.(\ref{eq:one_pt}). However, in real space, correlation functions
at two different angular scales are highly correlated, thus making error-analysis 
much more involved, even for all-sky coverage.

Two of the four three-to-one kurt-spectra $K_\ell^{B,2}$ and $K_\ell^{B,3}$ cannot 
be constructed using
(cubic) combinations of scalar fields, as they involve gradients $\nabla\myf$, and their co-ordinate independent constructions involve spinorial harmonics.
From the point of view of construction of MFs $K_\ell^{A,i}$ and $K_\ell^{B,i}$ carry equivalent
information. Also, as the spectra $K_\ell^{A,i}$ are not expressible as 
normalized lensing power spectrum $\myC_\ell^{\phi\phi}$, they are less appealing for
numerical implementation.
\section{Estimators, Mask, Noise and Covariances}
\label{sec:estim}
We have so far derived results for an ideal noise-free all-sky
survey. In reality partial sky coverage and
instrumental noise (possibly inhomogeneous) need to be dealt with.
Partial sky coverage introduces mode-mode coupling in the
harmonic domain in such a way that individual masked harmonics
become linear combinations of all-sky harmonics. The coefficients
for this linear transformation depend on specific choice of mask
through its own harmonic coefficients. Based on the 
pseudo-${\cal C}_{\ell}$ (PCL) method devised in ref.\cite{Hiv} for
power spectrum analysis,
{\em unbiased} estimators for skew-spectra and kurt-spectra 
were later developed in  ref.\cite{Mun11} and ref.\cite{Mu_kurt10}
that can handle realistic data.
\subsection{Estimators}
Consider two generic fields $U(\oh)$ and $V(\oh)$ and denote their
harmonic decompositions in the presence of a generic mask $w(\oh)$ as
$\tilde U_{\ell m}$ and $\tilde V_{\ell m}$. The fields $U$
and $V$ may correspond to any of the fields we have considered
above and  the harmonics $\tilde U_{\ell m}$ and $\tilde V_{lm}$ will correspond to
any of the harmonics listed in Eq.(\ref{eq:k0})- Eq.(\ref{eq:k4}) i.e.,
$[\myf^2]_{\ell m}$, \; $[\nabla \myf\cdot \nabla \myf]_{\ell m}$ and
$[\nabla^2 \myf]_{\ell m}$. The {\em pseudo-harmonics} of the masked fields are linear combinations
of the ordinary harmonics:
\ben
&& \tilde U_{LM} = \int ~d\oh~ Y^*_{LM}(\oh)~ [w(\oh)~ U(\oh)]; \\
&& \tilde U_{LM} = \sum_{\ell_im_i} (-1)^m~I_{L\ell_1\ell_2} \left ( \begin{array}{ c c c }
     \ell_1 & \ell_2 & L \\
     m_1 & m_2 & -M
  \end{array} \right) w_{\ell_1m_1} U_{\ell_2m_2};\\
&& w_{lm} = \int~d\oh~Y^{*}_{\ell m}(\oh)~w(\oh). 
\een \n
The construction of the {\em pseudo} kurt-spectra $\tilde K_\ell^{U,V}$ simply involves cross-correlating the relevant pseudo harmonics $\tilde U$ and $\tilde V$:
\ben && \tilde K_L^{U,V}\nfw
= {1 \over \Xi_{L}} \Re\left [ \sum_{m} \tilde U_{LM} \tilde V^*_{LM} \right ]; \quad
\tilde K_L^{U,V}\nfw = \sum_{\ell'} M_{L\ell'} K_{\ell^{\prime}}^{U,V}\nfw. 
\een
The mixing matrix $M$ is a function of the power spectrum $w_{\ell}$ of the mask $w(\oh)$:
\ben
&& M_{LL'} = {1 \over \Xi_{L}}\sum_{\ell} I^2_{LL'\ell} |w_{\ell}|^2;
\quad
\hat K_L^{U,V}\nfw = \sum_{\ell'} [M^{-1}]_{L\ell'} \tilde K_{\ell'}^{U,V}\nfw; \\
&& w_{\ell m} = \int w(\oh) Y^*_{\ell m}(\oh)d\oh; \quad\quad w_\ell = {1 \over \Xi_{\ell} }\sum_{m}w_{\ell m}w^*_{\ell m}.
\een
The estimator $\hat K_L^{U,V}\nfw$ constructed from 
pseudo-$\myC_\ell$s is unbiased as  $\la \hat K_L^{U,V}\nfw \ra =  K_L^{U,V}\nfw$.
The scatter from the ensemble mean $\delta \hat K_L^{U,V}$  and its covariance 
$\la\delta \hat K_L^{U,V}\delta \hat K_L^{U',V'}\ra$ can be computed
using the expressions given below:
\ben
&& \langle \hat K_L^{U,V}\nfw \rangle = K_L^{U,V}\nfw;  \quad
 \delta K_L^{U,V}\nfw =  \hat K_{L}^{U,V}\nfw -
\la K_L^{U,V}\nfw \ra ;\\
&& \la \delta \hat K_L^{U,V}\nfw \delta \hat K_{L'}^{U,V}\nfw \ra =
\sum_{LL'}M^{-1}_{L\ell} \la \delta \tilde K_{\ell}^{U,V}\nfw \delta \tilde K_{\ell'}^{U,V}\nfw \ra M^{-1}_{\ell'L'};\\
&& \quad\quad \left \{ U,V \right \} \in \left
\{\myf,\myf^2, (\nabla \myf\cdot \nabla \myf), \nabla^2 \myf
\right\}. \label{eq:auto_cov} \een
For small sky-coverage the matrix $M_{\ell\ell'}$ is singular and broad binning 
in the $\ell$ space may be required before the inversion.
\begin{figure}
\vspace{1.25cm}
\begin{center}
{\epsfxsize=10. cm \epsfysize=5. cm 
{\epsfbox[50 407 389 585]{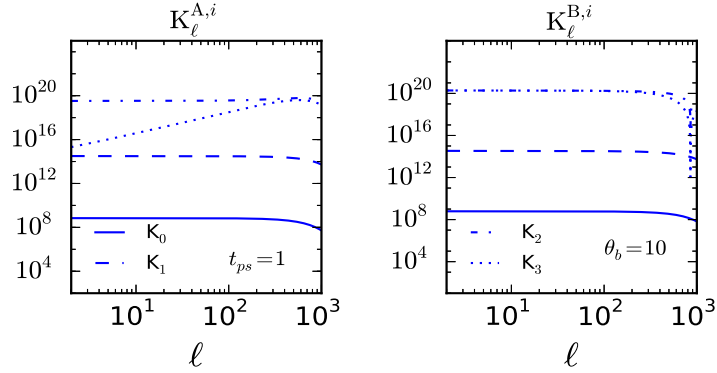}}}
\caption{The four kurtosis-spectra $K^{A,i}_{\ell}$ as defined in Eq.(\ref{eq:kurt_spectra2a})
for unresolved point sources are shown in the left panel. Corresponding results for$K^{A,i}_{\ell}$ are defined 
in Eq.(\ref{eq:kurt_spectra2b}). The expression for the trispectrum of point-sources is given by Eq.(\ref{eq:unresolved}).
In each panel solid, dashed, dot-dashed and dotted lines correspond to $\rm K^{(0)}_\ell$, $\rm K^{(1)}_\ell$, $\rm K^{(2)}_\ell$ 
and $\rm K^{(3)}_\ell$ respectively. We assume $\theta_0=10^{\prime}$ and an all-sky coverage is assumed.}
\label{fig:pt}
\end{center}
\end{figure}   
\subsection{Error Covariance}
The derivation of the
covariance depends on a Gaussian approximation i.e. we ignore higher-order non-Gaussianity in the fields.
$\myC_\ell$ is the ordinary CMB power spectra it also includes the effect of instrumental noise and beam $\myC_\ell^t\nfw = \myC^S_\ell b_\ell^2\nfw + n_\ell$. Such an approximation is suitable for
noise-dominated surveys. Moreover,
for a survey with homogeneous noise, we can write $n_\ell = \Omega_p \sigma_N^2$ where
$\Omega_p$ is the pixel area and $\sigma_N$ is the noise r.m.s. 
The relations listed below, constructed using Wick's theorem, will be useful in
derivation of scatter for our estimators:
\ben
&& \la \delta K_L^{U,V}\nfw \delta K_{L}^{U,V}\nfw \ra_c = {1 \over \Xi_{L}}
\left [ K_L^{U,U}\nfw \; K_L^{V,V}\nfw + [K_L^{U,V}\nfw]^2 \right ];\label{eq:err1} \\
&& \la \delta K_L^{U_1,V_1}\nfw \delta K_{L}^{U_2,V_2}\nfw \ra_c =  
{1 \over \Xi_{L}}  \left [ K_L^{U_1,U_2}\nfw\;K_L^{V_1,V_2}\nfw +K_L^{U_1,V_2}\nfw\;K_L^{U_2,V_1}\nfw \right ].
\label{eq:err2} \een 
The explicit  expressions for the scatter and covariance that we will require are listed below: 
\ben
&& \la {\delta K^{A,0}_{L} \delta K^{A,0}_{L'}} \ra \equiv 
{\delta_{LL^{\prime}}} {2f_{\rm sky}^{-1}\over \Xi_{L}}A_0^2 \Big[ K_L^{ {\Theta^2,\Theta^2} }\Big]^2; \label{eq:var1}\\
&& \la {\delta K^{A,1}_{L}\delta K^{A,1}_{L^{\prime}}} \ra \equiv  
\delta_{LL^{\prime}}{f_{\rm sky}^{-1}\over \Xi_{L}}A_1^2\left [
 K_L^{\Theta^2,\Theta^2 } K_L^{\Theta\nabla^2\Theta,\Theta\nabla^2\Theta}+  
[K_L^{\Theta^2,\Theta\nabla^2\Theta}]^2 \right ]; \label{eq:var2} \\
&& \la \delta K^{A,2}_L\delta K^{A,2}_{L^{\prime}} \ra_c=
\delta_{LL^{\prime}}{f_{\rm sky}^{-1}\over \Xi_{L}} A_2^2
 \Big [4\,K_{L}^{\Theta\nabla^2\Theta,\Theta\nabla^2\Theta}K^{\nabla\Theta\nabla\Theta,\nabla\Theta\cdot\nabla\Theta}_{L} + 
4\,[K_L^{\Theta\nabla^2\Theta,\nabla\Theta\cdot\nabla\Theta}]^2  \nn  \label{eq:var3}\\
&& \hspace{3cm}+4\,K^{\Theta\nabla^2\Theta,\nabla\Theta\cdot\nabla\Theta}_{L}K_{L}^{\nabla\Theta\cdot\nabla\Theta,\nabla\Theta\cdot\nabla\Theta}+ 2\,[K_L^{\nabla\Theta\cdot\nabla\Theta,\nabla\Theta\cdot\nabla\Theta}]^2 \Big ];\\
&&  \la {\delta K^{A,3}_{L}\delta K^{A,3}_{L^{\prime}}} \ra \equiv 
\delta_{LL^{\prime}}{2f_{\rm sky}^{-1}\over \Xi_{L}}{A_3^2} \Big [K_L^{\nabla\Theta\cdot\nabla\Theta,
\nabla\Theta\cdot\nabla\Theta} \Big ]^2.
\label{eq:var4} \een
Note that $\hat K_\ell^{A,2}$ gets contributions from two separate terms.
Here $f_{sky}$ is the fraction of sky-coverage for the survey under consideration.
Notice that unlike the skew-spectra and their generalisations introduced in ref.\citep{modgrav14}
kurtosis-spectra are not correlated in the limiting case of all-sky coverage.
In this respect they are similar to the ordinary power spectrum.
The individual expressions depends on the power spectrum: 
\ben
&& K_L^{\Theta^2, \Theta^2}= {2}{1 \over \Xi_{L}}\sum_{\{\ell_i\}=2}^{\ell_{max}} 
I^2_{\ell_1\ell_2L} \myC^t_{\ell_1}\myC^t_{\ell_2} ; \quad \label{eq:t1}\\
&& K^{\Theta\nabla^2\Theta ,\Theta\nabla^2\Theta}_L = {1 \over \Xi_L}\, {}\sum_{\{\ell_i\}=2}^{\ell_{max}} I^2_{\ell_1\ell_2L} 
\left [ \Pi_{\ell_1} + \Pi_{\ell_2} \right ]\, \Pi_{\ell_2} \myC^t_{\ell_1}\, \myC^t_{\ell_2}.\label{eq:t2}\\
&& K^{\Theta^2 ,\Theta\nabla^2\Theta}_L = {-}{1 \over \Xi_L}\sum_{\{\ell_i \}=2}^{\ell_{max}}  I^2_{\ell_1\ell_2L}\left [\Pi_{\ell_1}+\Pi_{\ell_2}\right ]\myC^t_{\ell_1}\myC^t_{\ell_2}\label{eq:t3}\\
&& K_L^{\nabla\Theta\cdot\nabla\Theta,\nabla\Theta\cdot\nabla\Theta}=
{}{1 \over 2}{1 \over \Xi_L}\sum_{\{\ell_i\}=2}^{\ell_{max}} I^2_{\ell_1\ell_2L}
\; [{\Pi}_{\ell_1}+ {\Pi}_{\ell_2}- {\Pi_{L}}]^2 \; {\cal C}^{t}_{\ell_1}{\cal C}^{t}_{\ell_2}\label{eq:t4}\\
&& K_L^{\Theta\nabla^2\Theta,\nabla\Theta\cdot\nabla\Theta} =-{1 \over \Xi_L}\sum_{\{\ell_i\}=2}^{\ell_{max}}
{I^2_{\ell_1\ell_2L}}
\; [{\Pi_{\ell_1}}+{\Pi}_{\ell_2}-{\Pi}_{L}]\;{\Pi_{\ell_2}}\;  {\cal C}^{t}_{\ell_1}{\cal C}^{t}_{\ell_2} \label{eq:t5};\\ 
&& K_L^{\Theta^2,\nabla\Theta\cdot\nabla\Theta} = {1\over \Xi_L}\sum_{\{\ell_i\}=2}^{\ell_{max}}
{I^2_{\ell_1\ell_2L}}
\; [{\Pi_{\ell_1}}+{\Pi}_{\ell_2}-{\Pi}_{L}]\;  {\cal C}^{t}_{\ell_1}{\cal C}^{t}_{\ell_2} \label{eq:t6},
\een
where ${\cal C}^t_{\ell}$ is the total power spectrum, including contributions from detector noise.
The expressions are symmetric under exchange of indices that are summed over i.e. $\ell_1$ and $\ell_2$
and we can restrict the summation to the upper triangular matrix 
$\sum^{\ell_{max}}_{\{\ell_i\}}= 2\sum^{\ell_{max}}_{\ell_1=2}\sum^{\ell_{max}}_{\ell_2 =\ell_1}$.
We have included the expression in Eq.(\ref{eq:t6}) will be required for the calculation
of covariances. 
The signal-to-noise $[\rm S/ \rm N]$ for individual modes for a given spectrum
on the other hand can be expressed as:
\ben
&& [{\rm S/\rm N}]^{A,(i)}_L(\fw) =  \sqrt{\la [K_L^{A,(i)}(\fw)]^2\ra /\la [\delta K_L^{A,(i)}(\fw)]^2\ra }\quad\quad i \in \{0,1,2\}.
\label{eq:s2n_def}
\een
In our estimates of scatter we neglect contributions from terms describing higher-order non-Gaussianity such as the trispectrum.
Thus, our results provide accurate results in the noise-dominated regime. For high sensitivity experiments
Monte-Carlo simulation is the only way to evaluate the scatter. 
Also, we have assumed a uniform white noise, whereas in real experiments the noise will be non-uniform. Such complications can only be dealt  with by running simulations.
The parameters for a few ongoing and planned experiments
are tabulated in Table \ref{tabular:tab1}, and the corresponding cumulative S/N for various estimators
are listed in Table \ref{tabular:tab_s2n}.

The estimators and the scatter are not independent. To compute the cross-correlation in scatter
we will need the following expressions:
\ben
&& \la\delta K^{A,0}_{L}\delta K^{A,1}_{L^{\prime}} \ra_c \equiv 
\delta_{LL^{\prime}}{2 \fsky \over \Xi_{L}}A_0 A_1 \left[
K_L^{\Theta^2,\Theta^2} K_L^{\Theta^2,\Theta\nabla^2\Theta}\right ]; \label{eq:r1}\\
&& \la\delta K^{A,0}_{L}\delta K^{A,2}_{L^{\prime}} \ra_c \equiv 
\delta_{LL^{\prime}}{2 \fsky \over \Xi_{L}} A_0 A_2 \left[
2 K_L^{\Theta^2,\Theta\nabla^2\Theta}K_L^{\Theta^2,\nabla\Theta\cdot\nabla\Theta}
+[ K_L^{\Theta^2,\nabla\Theta\cdot\nabla\Theta}]^2 \right ]; \label{eq:r2}\\
&& \la\delta K^{A,0}_{L}\delta K^{A,3}_{L^{\prime}} \ra_c \equiv 
\delta_{LL^{\prime}}{2 \fsky \over \Xi_{L}} \; A_0 A_3 \left [K_L^{\Theta^2,\nabla\Theta\cdot\nabla\Theta}\right ]^2; 
\label{eq:r3}\\
&&  \la \delta K_L^{A,1}\delta K_{L^{\prime}}^{A,2} \ra_c= \delta_{LL^{\prime}}{\fsky \over \Xi_{L}} \; A_1 A_2 
\Big[ 2\, K_{L}^{\Theta^2, \nabla\Theta\cdot\nabla\Theta}
K_{L}^{\Theta\nabla^2\Theta, \Theta\nabla^2\Theta} \nn\\
&& \hspace{2cm} + 2\,K_{L}^{\Theta\nabla^2\Theta,\nabla\Theta\cdot\nabla\Theta}\,
K_{L}^{\Theta^2, \Theta\nabla^2\Theta} +
K^{\Theta^2,\nabla\Theta\cdot\nabla\Theta}_L K^{\Theta\nabla^2\Theta,\nabla\Theta\cdot\nabla\Theta}_L  \Big ]; \label{eq:r4}\\
&& \la \delta K_L^{A,1}\delta K_{L^{\prime}}^{A,3} \ra_c=
\delta_{LL^{\prime}}{2 \fsky\over \Xi_{L}}A_1 A_3 \left [K_L^{\Theta^2,\nabla\Theta\cdot\nabla\Theta} 
K_L^{\Theta\nabla^2\Theta,\nabla\Theta\cdot\nabla\Theta} \right ]; \label{eq:r5} \\
&& \la \delta K_L^{A,2}\delta K_{L^{\prime}}^{A,3} \ra_c\equiv 
\delta_{LL^{\prime}}{2 \fsky \over \Xi_{L}} A_2 A_3  
\Big[2 K^{\Theta\nabla^2\Theta,\nabla\Theta\cdot\nabla\Theta}_{L} K^{\nabla\Theta\cdot\nabla\Theta,\nabla\Theta\cdot\nabla\Theta}_{L} \nn\\
&& \hspace{5cm}+ [K^{\nabla\Theta\cdot\nabla\Theta,\nabla\Theta\cdot\nabla\Theta}_{L}]^2 \Big ].
\label{eq:r6}
\een
The terms that appear in Eq.(\ref{eq:r1})-Eq.(\ref{eq:r6}) can all be expressed in terms of quantities 
defined in Eq.(\ref{eq:t1})-Eq.(\ref{eq:t6}).
Notice that different harmonic modes of different estimators are uncorrelated in the all-sky limit.
These expressions are used to compute the cross-correlation coefficient among
various spectra which are defined below:
\ben
r^{ij}_L(\fw) = \la \delta K_L^{A,(i)}(\fw) \delta K_L^{A,(j)}(\fw)\ra /
\sqrt{ \la [\delta K_L^{A,(i)}(\fw)]^2\ra \la [\delta K_L^{A,(j)}(\fw)]^2 \ra};
i,j \in \{0,1,2\}.
\label{eq:cross_def}
\een
Throughout we have ignored mode-mode coupling. The coefficients of cross-correlation  $r_{ij}$ are
independent of the sky-coverage $f_{\rm sky}$ and normalisation coefficients $A_i$. Cross-correlation
of generalised spectra $K^{A,i}_\ell$ and $K^{B,i}_\ell$ defined in 
Eq.(\ref{eq:kurt_spectra2a})-Eq.(\ref{eq:kurt_spectra2b}) will vanish in the Gasussian limit
as they will involve odd-order leading terms.

Notice that in our computation of error estimates we have ignored the error in the power spectrum $\myC_\ell$,
and assumed that the variances $\sigma_0^2$ and $\sigma_1^2$ are known exactly. 

\subsection{Computation of $\chi^2$}
The departure between General Relativity (GR) and a modified gravity (MG) model, from various two-to-two estimators, can be quantified by: 
\ben
\chi^2_{\phi\phi} = \sum_{ij}\sum_{\ell\ell'} \delta {\cal C}^{\phi\phi,(i)}_{\ell} \;{\mathbb C}^{\phi\phi}_{ij,\ell\ell^{\prime}}\; \delta{\cal C}^{\phi\phi, (j)}_{\ell^{\prime}}; \quad\quad
{\delta {\cal C}}^{\phi\phi,(i)}_{\ell} = {\cal C}^{\phi\phi,(i)}_{\ell}|_{\rm MG} - {\cal C}^{\phi\phi,(i)}_{\ell}|_{\rm GR}
\een
We notice from Eq.(\ref{eq:no0})-Eq.(\ref{eq:no3}) that we can construct an estimator for ${\delta {\cal C}}^{\phi\phi,(i)}_{\ell}$ 
from each  $K^{A,(i)}_{\ell}$:
\ben
{{\cal C}}^{\phi\phi,(i)}_{\ell} =  [N^{(i)}_{\ell}]^{-1}K^{A,(i)}_{\ell}; \quad  {\mathbb C}^{\phi\phi}_{ij,\ell\ell^{\prime}} = 
[N^{(i)}_{\ell}]^{-1}[N^{(j)}_{\ell^{\prime}}]^{-1}\la{\delta {K}}^{A,(i)}_{\ell}{\delta {K}}^{A,(j)}_{\ell^{\prime}}\ra
\een 
Using these relations we can directly evaluate the $\chi^2$ using the covariance of $K^{A,(i)}_\ell$ (which we will denote as ${\mathbb C}^{A}_{ij,\ell\ell^{\prime}}$)
presented in Eq.(\ref{eq:var1})-Eq.(\ref{eq:var4}) and
Eq.(\ref{eq:r1})-Eq.(\ref{eq:r6}). :
\ben
&& \chi^2_{\phi\phi} = \sum_{ij}\sum_{\ell} \delta {K}^{A,i}_{\ell} \;[{\mathbb C}^{A}]^{-1}_{ij,\ell\ell}\; \delta{K}^{A,j}_{\ell^{\prime}};
\label{eq:chi2A} \\
&& {\mathbb C}^{A}_{ij,\ell\ell} = \la {\delta K}^{A,i}_{\ell}{\delta K}^{A,i}_{\ell}\ra_c; \quad
{\delta {K}}^{A,i}_{\ell} = {K}^{A,i}_{\ell}|_{\rm MG} - {K}^{A,i}_{\ell}|_{\rm GR} \label{eq:chi2B}
\een
We have used the fact that ${\mathbb C}^{A, i}$ is diagonal for an all-sky experiment. We will specialise this expression for $K^{A,3}_{\ell}$
as the (S/N) is considerably higher for this estimator and the other estimators have significant correlation with $K^{A,3}_{\ell}$.

\section{Results and Discussion}
\label{sec:disc}
\begin{figure}
\vspace{1.25cm}
\begin{center}
{\epsfxsize=10. cm \epsfysize=4.5 cm 
{\epsfbox[44 414 383 585]{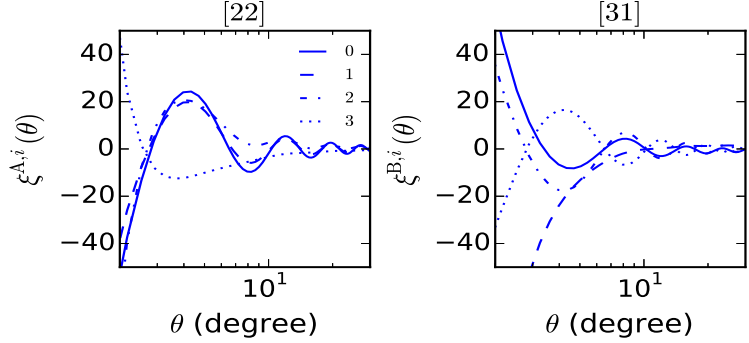}}}
\label{fig:corr}
\caption{The correlation functions $\xi^{{\rm A},i}(\theta)$ (left-panel) and $\xi^{{\rm B},i}(\theta)$ (right-panel)
defined respectively in Eq.(\ref{eq:corrA}) and Eq.(\ref{eq:corrB})  are shown as a function of $\theta_{12}$. A $\Lambda$CDM cosmology
and GR is assumed. 
The solid, dashed, dot-dashed and dotted lines in left panel correspond to $\rm (A,0)\times 10^{-5}$, $\rm (A,1)\times 10^1$, $\rm (A,2)\times 10^7$ 
and $\rm (A,3)\times 10^7$ respectively. In the right panel they correspond to $\rm (B,0)\times 10^{-5}$, $\rm (B,1)\times 10^{1}$, 
$\rm (B,2)\times 10^{7}$ and $\rm (B,3)\times 10^{7}$. We have assumed the smoothing angular scale as $\theta_0=10^{\prime}$ and $\ell_{max}=10^3$.}  
\end{center}
\end{figure}

\begin{table}
\begin{center}
\caption{Cumulative ${\rm [S/N]}/f_{\rm sky}$ for various surveys}
\vspace{0.5cm}
\begin{tabular}{| c | c | c | c | c | }
  \hline
  & K$^{A,3}_\ell$ & K$^{A,2}_\ell$ & K$^{A,1}_\ell$  & K$^{A,0}_\ell$ \\
  \hline
  \rowcolor[gray]{0.8} ACTPol &  1400 & 570 & 6.0& 1.0 \\
 \hline
COrE$^{+}$ & 2300 & 720 & 10.0  & 3.0 \\
\hline
\rowcolor[gray]{0.8} Planck & 4.0 & 2.0 & $8\times 10^{-2}$ & $2\times 10^{-2}$\\
\hline
\end{tabular}
\label{tabular:tab_s2n}
\end{center}
\end{table}
\begin{table}
\begin{center}
\caption{ACTPol $\chi$ for various models}
\vspace{0.5cm}
\begin{tabular}{| c | c | c | c | c |c | c| }
  \hline
   & H2 &  H3 & Q & EFT$_1$  & EFT$_2$ & B$_0$ \\
  \hline
  \rowcolor[gray]{0.8} K$_3$ &  65 & 57 & 17 & 20 & 30 & 18 \\
 \hline
$\rm K_2$ & 1.5 & 1.0 & 3.9 & 5.0 & 7.0 & 4.0\\
\hline
 \rowcolor[gray]{0.8} $\rm K_1$ & 0.15 & 0.15 & 0.04 & 0.02 & 0.06 & 0.05\\
\hline
 $\rm K_0$ & 0.04 & 0.04 & 0.01 & 0.006 & 0.02 & 0.01\\
\hline
\end{tabular}
\label{tabular:actpol}
\end{center}
\end{table}

\begin{table}
\begin{center}
\caption{COrE$^{+}$ $\chi$ for various models}
\vspace{0.5cm}
\begin{tabular}{| c | c | c | c | c |c | c| }
  \hline
   & H2 &  H3 & Q & EFT$_1$  & EFT$_2$ & B$_0$ \\
  \hline
  \rowcolor[gray]{0.8} K$_3$ &  94 & 87 & 25 & 26 & 43 & 30 \\
 \hline
$\rm K_2$ & 17 & 15 & 4.3 & 5.3 & 7.8 & 4.8\\
\hline
 \rowcolor[gray]{0.8} $\rm K_1$ & 0.4 & 0.4 & 0.1 & 0.05 & 0.18 & 0.13\\
\hline
 $\rm K_0$ & 0.16 & 0.16 & 0.04 & 0.02 & 0.07 & 0.05\\
\hline
\end{tabular}
\label{tabular:core}
\end{center}
\end{table}

\begin{table}
\begin{center}
\caption{Planck $\chi$ for various models}
\vspace{0.5cm}
\begin{tabular}{| c | c | c | c | c |c | c| }
  \hline
   & H2 &  H3 & Q & EFT$_1$  & EFT$_2$ & B$_0$ \\
  \hline
  \rowcolor[gray]{0.8} K$_3$ &  1.7 & 1.5 & 0.4 & 0.5 & 0.8 & 0.5 \\
 \hline
$\rm K_2$ & .40 & .36 & 0.1 & 0.01 & 0.19 & 0.1 \\
\hline
 \rowcolor[gray]{0.8} $\rm K_1$ & 0.02 & 0.02 & 0.006 & 0.004 & 0.01 & 0.008\\
\hline
 $\rm K_0$ & 0.008 & 0.008 & 0.002 & 0.001 & 0.003 & 0.003\\
\hline
\end{tabular}
\label{tabular:planck}
\end{center}
\end{table}

\begin{figure}
\vspace{1.25cm}
\begin{center}
{\epsfxsize=14. cm \epsfysize=4.5 cm 
{\epsfbox[38 417 545 585]{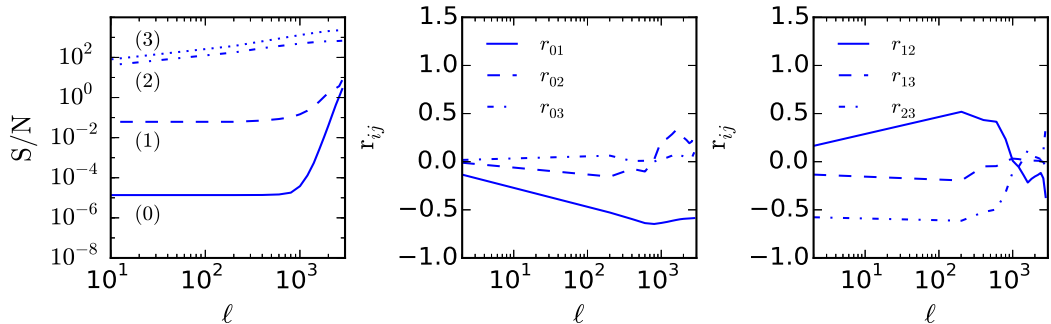}}}
\caption{The {\em cumulative} sum of the $\rm S/N$ is plotted as a function of $\ell$. The solid, dashed, dot-dashed
and dotted lines correspond to $K^{(0)}_\ell$, $K^{(1)}_\ell$, $K^{(2)}_\ell$ and $K^{(3)}_\ell$. The $\rm S/N$ is
defined in Eq.(\ref{eq:s2n_def}).
The individual modes of a given kurt-spectra are uncorrelated. However, for a given $\ell$
the estimates of various kurt-spctra are correlated. The correlation coefficients defined in Eq.(\ref{eq:cross_def}) are
plotted as a function of $\ell$ in middle and right panel. The middle panel shows $r_{01}$ (solid-lines),
$r_{02}$ (dashed-lines) and $r_{03}$ (dot-dashed lines). The right panel shows $r_{23}$ (solid-lines), 
$r_{24}$ (dashed-lines) and $r_{34}$ (dot-dashed lines) respectively.
We have taken $\ell_{max}=3000$ and the noise and beam correspond to that of the ACTPol survey. (see text for more details).}  
\label{fig:s2n_cross}
\end{center}
\end{figure}

\begin{enumerate}
\item
The kurt-spectra we have defined depend on the lensing power spectra and the CMB temperature power spectra. 
In Fig.\ref{fig:phiphi} we show the $\cal C^{\phi\phi}_{\ell}$ as a function of $\ell$ for various
MG theories. In Fig.\ref{fig:isw} the low-mulipole sector of the temperature
power-spectrum ${\cal C}_{\ell}$ is shown. 
\item
{\bf The two-to-two and three-to-one estimators:}
The two-to-two and three estimators for various theories of gravity are shown in Fig.\ref{fig:A31} and Fig.\ref{fig:B22}.
The expressions in Eq.(\ref{eq:kurt_spectra2a}) and Eq.(\ref{eq:kurt_spectra2b}) define these two spectra.
In our construction of these we have used the approximation $\rm T=P$, see e.g. Eq.(\ref{Total_Tri}), thereby ignoring the terms that involve
computation of $6j$ coefficients. This approximation is commonly used in the literature and produces
results which are reasonably accurate \citep{Hu01,Smidt10}. This simplified our analytical results. Use of this approximation
makes all two-to-two estimators directly proportional to the lensing potential power-spectra ${\cal C}^{\phi\phi}_{\ell}$.
The normalisation coefficients depend on the harmonics $\ell$ and can be computed once the background cosmology
is known. For the two-to-two spectra the individual $\ell$ modes are uncorrelated for all-sky surveys,
which makes computation of statistics such as the $\chi^2$ statistic rather trivial. 
It depends only on the fiducial temperature power spectra. The two-to-two kurt-spectra for 
the unresolved point sources can be constructed equally easily. The correlation functions that we 
can define from these spectra  Eq.(\ref{eq:corrA}) and Eq.(\ref{eq:corrB})  are shown in  Fig.\ref{fig:corr}.
To compute the correlation functions we have included harmonics up to $\ell_{max}=10^3$ and a FWHM of $\theta_0=10^{\prime}$.

The three-to-one spectra defined in Eq.(\ref{eq:b0})-Eq.(\ref{eq:b3}) on the other hand
depend on the lensing spectra through a convolution, which makes their interpretation
complicated. Indeed, unlike the two-to-two estimators, the error-covariance of three-to-one spectra 
includes off-diagonal terms, even in the absence of any mask or non-uniform noise.
In fact, it can be shown that all higher-order spectra at even order which are constructed by
cross-correlating same combination of fields will have diagonal covariance matrix
e.g. the ordinary power spectra which is one-to-one. At sixth order, the
three-to-three spectra share the property of the two-to-two spectra, in having diagonal covariance, with an $f_{\rm sky}$
prescription being used to take into account the mode-mode coupling.
However, any other spectra
such as the two-to-one skew-spectra or the three-to-one kurt spectra will exhibit
off-diagonal elements, as they are not an auto-spectra of a quadratic $[\Theta^2]$ or cubic $[\Theta^3]$ combination.

The implementation of three-to-one spectra is also difficult, since to decompose a cubic combination
involving partial derivatives we have to use spinorial spherical harmonics. 
\item
{\bf The variance and cross-correlation:} 
The variance of the estimators is presented in Fig.\ref{fig:s2n_cross}. Of all the estimators we have 
studied the estimator $K^{A,3}_\ell$ has the maximum (S/N) followed by $K^{A,2}_{\ell}$. The (S/N) is typically high for $\ell = 10^2-10^3$.
The other two estimators lack the (S/N) needed to be useful.
The expressions in Eq.(\ref{eq:var1})-Eq.(\ref{eq:var4}) gives the expressions for the variance in these estimators.
This is in qualitative agreement with ref.\citep{KCM02} where the ordinary kurtosis parameter was studied and was 
found to lack the value of (S/N) needed for detection even in a cosmic variance dominated survey.
Eq.(\ref{eq:r1})-Eq.(\ref{eq:r6}) defines the cross-correlations. To compute the scatter and correlation of our estimators
are defined in terms of quantities defined in the expressions in Eq.(\ref{eq:t1})-Eq.(\ref{eq:t6}). Computations of
these expressions are based on the assumption that the underlying CMB harmonics are Gaussian. All higher order
non-Gaussianities are ignored, and we have assumed that the noise is independent of pixel position to simplify our analytical results. 
It is indeed possible
to define more sophisticated estimators that work directly with the {\em Wiener-filtered} harmonics \cite{ducout13}
to improve the (S/N). For more accurate estimates of (S/N)
it is possible to employ simulation chain using software such as the Lenspix\footnote{http://cosmologist.info/lenspix/}
that can handle inhomogeneous noise.
The power-spectra ${\cal C}^{\phi\phi}_{\ell}$ we have used are based on linear theory. Numerical simulations and non-linear
modelling have been used to compute the nonlinear corrections to ${\cal C}^{\phi\phi}_{\ell}$ \citep{carbone13}. 
Such calculations
are lacking at present for many of the MG theories we have considered. Construction of our estimators
do not depend on the shape of $\cal C^{\phi\phi}_\ell$ and inclusion of non-linearity is unlikely to change
the qualitative results presented here. 
\item
{\bf MG theories and $\chi^2$}: In Table \ref{tabular:tab_s2n} we find that surveys such as ACTPol or CoRE$^{+}$\footnote{http://www.core-mission.org/} should be able to 
detect the lensing of the CMB in temperature maps with extremely high S/N. 
However, a few comments are in order, since we have used a simple model for covariance. In practice,
we will have to deal with inhomogeneous noise, the connected part of the covariance matrix and
any residuals from component separation. 
The $\chi^2_{\phi\phi}$ defined in Eq.(\ref{eq:chi2A})-Eq.(\ref{eq:chi2B}) 
for various survey configuration are displayed in Tables \ref{tabular:actpol},\ref{tabular:core},\ref{tabular:planck}.

However, even if we consider our estimates as no more than an order magnitude estimate    
they still are impressive. The models that we have considered are already rejected
by other observational data e.g. the $f(R)$ models.
For reference, we note that the designer $f(R)$ gravity, the constraint on its Compton wavelength parameter
 ($B_0$) from the latest Planck-2015 data are presented in ref.~\cite{Planck_MG}, $B_0<0.12~(0.04)$ at $95\%$ C.L. 
by using the compilation of temperature and low multipole polarization data, 
(the number in the parentheses is the one adding CMB lensing data). 

Our study suggests that the future lensing
data will further tighten the constraints and render them comparable to constraints from the
local tracers of large-scale structure of the Universe such as the weak-lensing surveys or
galaxy surveys. Also, we have focussed on MG theories but similar results are expected for neutrino mass hierarchy (Munshi et al. in prep.).

Inclusion of polarization data will further improve the result.
In addition to the surveys we have focussed there are many surveys that are 
being planned such as the EBEX\footnote{http://groups.physics.umn.edu/cosmology/ebex/}, 
Simons Array\footnote{http://cosmology.ucsd.edu/simonsarray.html} survey or the LiteBird\footnote{http://litebird.jp/eng/} surveys. 
\end{enumerate}
\section{Conclusions and Outlook}
\label{sec:conclu}
In this paper, we have studied how lensing of the CMB can change the topological
properties of temperature and polarization maps by using morphological descriptors such
as the Minkowski Functionals. A perturbative expansion links MFs
with the multispectra of the lensed maps. Recent studies have 
shown how statistics such as the skew-spectra \cite{MuHe10} can be 
valuable in reconstructing MFs of frequency cleaned maps at the level of 
lensing-secondary bispectrum, frequency-cleaned tSZ maps \cite{MSJC12},
maps from weak lensing surveys \cite{MWSC12}. The primary aim of this study
was to extend these results to include the lensing induced trispectrum in the analysis.
We have used the kurt-spectra introduced in ref.\cite{MCCHS11} for this purpose.

The shape of the kurt-spectra is a natural
diagnostics in distinguishing different sources of non-Gaussianity.
We construct a set of four kurt-spectra based estimators that are directly
proportional to the power spectrum $\myC_{\ell}^{\phi\phi}$ of the projected lensing-potential $\phi$,
which is a sensitive probe of the neutrino mass hierarchy and DE equation of state, and thus
provide a set of {\em sub-optimal} estimators for the {\em reconstruction} of lensing potential  power spectrum.
We have used them to study various MG theories.  
\begin{enumerate}
\item
{\bf Component Separation using non-Gaussianity: kurt-Spectra and topology of reionization:}
It was pointed out  in ref.\cite{RS07} that lensing (a gravitational secondary) and kSZ (a scattering secondary) 
share many interesting properties. Both of these secondaries lack any frequency information that can help them to separate
from primary CMB. These secondaries on the other hand are non-Gaussian and this
information can be used to separate them. The kurtosis is the leading order non-Gaussianity for both kSZ 
and lensing of CMB as all odd order contributions vanish for both these secondaries \cite{Castro04}.
While lensing is an important probe of gravitational physics, kSZ is an important probe of 
reionization history of the Universe. Reionization can in principle be
inhomogeneous. In ref.\cite{RS07} only the three-to-one estimator was used and
the non-Gaussian contribution to lensing were ignored. The systematic analysis
we have performed for both three-to-one and two-to-two estimators show that
simultaneous analysis of these two effects is possible, and using
the two set of estimators can separate the contributions without
imposing any further constraints.
The topological estimators which we have developed here Eq.(\ref{eq:kurt_spectra2a})-Eq.(\ref{eq:kurt_spectra2b})
or its real-space analogs Eq.(\ref{eq:corrA})-Eq.(\ref{eq:corrB})
can be employed to understand the topology of inhomogeneous reionization
using data from next generation of experiments (Munshi et al. in prep. 2016). 
We provide explicit expressions for the error-analysis of these spectra
which are completely generic and can also be used for kSZ \citep{Castro04}.
\item
{\bf Polarization and separation of gradient and curl modes:} 
In future the measurement of ${\cal C}^{\phi\phi}_{\ell}$ will be able to provide much tighter constraints on
cosmological parameters using not just the temperature but polarization data. In addition
to provide tighter constraints on neutrino masses, DE models and MG theories the ultimate goal
of CMB experiments will be to detect the primordial gravity wave signals through the measurement
of B-polarization. The signal however is confused with lensing effect. The lensing effect converts
the dominant E-polarization to $B$-modes. The estimators designed here can be used
to separate the signal from primordial gravitational waves and lensing of E-modes.
We incorporated only the
{\em gradient} modes in our results. However the {\em curl} mode that can be useful also for computing contribution from
gravitational wave, cosmic strings or primordial magnetic field \citep{pranjal} can easily be included in our results. 
Generalisation  of the results discussed here to polarization will be presented elsewhere.   
\item
{\bf Separation of $f_{\rm NL}$ and $g_{\rm NL}$: }
The optimized versions of two-to-two and three-to-one estimators have been
used to probe primordial non-Gaussianity beyond the lowest order i.e. to
separate contributions from $\tau_{\rm NL}$ and $g_{\rm NL}$ \cite{Smidt10}. 
Although current experimental results are consistent with null detection
of non-Gaussianiy, consistency check from future experiments can be
performed using the estimators defined here beyond the lowest order.

The estimators can be generalised to the study of primordial trispectra 
using CMB spectral distortions \citep{spectral}.
\end{enumerate}

The pseudo-${\cal C}_\ell$ formalism discussed above is
sub-optimal but extremely fast and its error covariance can be
computed analytically as we have shown.
These estimators are sub-optimal. However, in near future 
large fraction of the sky will be covered by experiments which will have
very low detector noise and small FWHM, thus optimality of estimators
may not be a crucial requirement in the future.  

\section{Acknowledgements}
\label{acknow}
DM and PC acknowledge support from the Science and Technology
Facilities Council (grant number ST/L000652/1). DM would like to thank 
Julien Peloton, Donough Regan, Antony Lewis and Joseph Smidt for useful discussions.
DM also acknowledges Patrick Valageas and Geraint Pratten for related collaborations.
BH is partially supported by the Programa Beatriu de Pin\'os and Dutch Foundation for Fundamental Research on Matter (FOM).

\bigskip
\bibliography{paper.bbl}
\appendix
\section{Explicit Expressions for the Kurtosis-Spectra}
\label{sec:appendA}
\subsection{Computation of disjoint or Gaussian contribution}
The Gaussian components of the estimators are as follows:

\ben
&& {G}_{L}^{A,0}= A_0{\cal G}_{L}^{(a)};  \label{eq:no}\\
&& {G}_{L}^{A,1}= -A_1 {\cal G}_{L}^{(b)}; \label{eq:no1G}\\
&& {G}_{L}^{A,2}= -A_2 \left [{\Pi}^2_{L}\, {\cal G}_{L}^{(a)} -{\cal G}_{L}^{(c)}  \right];\\ 
&& G_{L}^{A,3} = A_3
\left [{{\Pi}_{L}}^2 {\cal G}_{L}^{(a)} -2{{\Pi}_{L}} {\cal G}_{L}^{(b)} +  {\cal G}_{L}^{(c)}\right ];\label{eq:no2G}\\
&&{\cal G}_{L}^{(a,b,c)} = {1 \over 2\pi}\sum_{\ell_1\ell_2} \; Z^{(a,b,c)} \;I^2_{\ell_1 L\ell_2} \myC_{\ell_1}\myC_{\ell_2};\\
&& Z^{(a)}_{\ell_1\ell_2} =1; \quad  Z^{(b)}_{\ell_1\ell_2} = {\Pi}_{\ell_1} + {\Pi}_{\ell_2}; \quad  Z^{(c)}_{\ell_1\ell_2}= ({\Pi}_{\ell_1} + {\Pi}_{\ell_2})^2.
\label{eq:def_Z}
\een
The first term in each of these expressions denotes the monopole contribution $\ell=0$,
and the second term corresponds to $\ell\ne 0$. Notice these expressions depend on the total power spectrum of
both signal (beam-convolved) and noise (which is assumed Gaussian) i.e.,
$\myC^t_{\ell} = \myC_{\ell}^S b_{\ell}^2(\theta_0) + n_{\ell}$.  
The corresponding three-to-one estimators in the Gaussian limit are
\ben
&& G^{B,0}_L = B_0\, S^{(00)}_L\\
&& G^{B,1}_{L} =2B_1\, S^{(01)}_L\\
&& G^{B,2}_{L}= B_2\, [S^{(20)}_{L}-S^{(02)}_{L}] \\
&& G^{B,3}_{L}= B_3\, [S^{[20]}_L+S^{[02]}_L-2S^{[11]}_L],
\een
where we have introduced the quantities $S^{pq}_{\ell}$ to simplify the expressions:
\ben
S^{pq}_L= {{\cal C}_{L} \over \Xi_L} \sum_{\ell_1 \ell_2}{{\cal C}_{\ell_1}\over \Xi_{\ell_2}} \Pi^p_{L}(\Pi_{\ell_1}+\Pi_{L})^q {\cal C}_{\ell_1}
I^2_{\ell_1L\ell_2}.
\een
If we use the same normalisation as the two-to-two estimators we have $A_i=B_i$.
\subsection{Lensing induced two-to-two Kurtosis-Spectra  $K_\ell^{A,(i)}$}
\label{sec:ka_ng}
We will specialise the kurt-spectra we have derived in the text of the paper for the special case of lensing-induced non-Gaussianity.
All of the two-to-two spectra or equivalently the $K_{\ell}^{A,i}$ estimators we have studied can be expressed as a product of the lensing power spectrum  $\myC_{\ell}^{\phi\phi}$,
times an $\ell$-dependent normalisation:
\ben
&& K_{L}^{A,0}= A_0\,{\cal C}^{\phi\phi}_{L}{1 \over \Xi^2_{L}} [{\cal E}_{L}^{(0)}]^2; \label{eq:no0}\quad\\
&& K_{L}^{A,1}= A_1 \, {\cal C}^{\phi\phi}_{L}{1 \over \Xi^2_{L}} {\cal E}_{L}^{(0)}\,{\cal E}_{L}^{(1)}; \label{eq:no1} \\
&& K_{L}^{A,2}= A_2\,{\cal C}^{\phi\phi}_{L} {1 \over \Xi^2_{L}}\left [{\Pi^2_{L}}[{\cal E}_{L}^{(0)}]^2 - [{\cal E}_{L}^{(1)}]^2\right];\quad  \label{eq:no2} \\
&& K_{L}^{A,3} = A_3 \,{\cal C}^{\phi\phi}_{L}{1 \over \Xi^2_{L}}\, [{\cal E}_{L}^{(1)} - {\Pi_{L}}{\cal E}_{L}^{(0)}]^2. \label{eq:no3}
\een
The factors ${\cal E}_L^{(0)}$ and ${\cal E}_L^{(1)}$ depend only on the power spectrum of the unlensed CMB sky $\myC_{\ell}$
and are defined below :
\ben
&& {\cal E}_{L}^{(0)} \equiv \sum_{\ell_1\ell_2}f_{\ell_1 L \ell_2}I_{\ell_1 L \ell_2}; \quad
{\cal E}^{(1)}_{L} \equiv \sum_{\ell_1\ell_2}(\Pi_{\ell_1}+\Pi_{\ell_2})f_{\ell_1L\ell_2}I_{\ell_1L\ell_2}.
\label{eq:E_matrix}
\een
For point sources we can use the expressions Eq.(\ref{eq:no0})-Eq.(\ref{eq:no3}) by redefining the ${\cal E}$ matrices:
\ben
&& {\cal E}_{L}^{{\rm ps},(0)} \equiv \sum_{\ell_1\ell_2}I^2_{\ell_1 L \ell_2}; \quad
{\cal E}^{{\rm ps},(1)}_{L} \equiv \sum_{\ell_1\ell_2}(\Pi_{\ell_1}+\Pi_{\ell_2})I^2_{\ell_1 L \ell_2}.
\label{eq:E_matrix_ps}
\een
The normalisation coefficients $A_i$ will remain unchanged but the ${\cal C}^{\phi\phi}_{\ell}$ will have to
be replaced with the amplitude for the unresolved point source trispectrum $t_{\rm ps}$ introduced in Eq.(\ref{eq:unresolved}).  
\subsection{Lensing induced three-to-one Kurtosis-Spectra  $K_\ell^{B,(i)}$}
\n
In this section we will compute the three-to-one term from lensing. 
The construction of the estimator follows exactly same procedure as for the two-to-two estimator. We list the expressions below :
\ben
&& K_L^{B,0} =  A_0\, {1 \over \Xi_{L}}\sum_\ell {\myC_\ell^{\phi\phi}\over \Xi_{\ell}} {\cal D}^{(0)}_{\ell L} {\cal E}^{(0)}_{\ell}; 
\label{eq:b0}\\ 
&& K_L^{B,1} =  A_1\, {1 \over \Xi_{L}}\sum_L {\myC_\ell^{\phi\phi}\over \Xi_{\ell}}  
\left [ {\cal D}^{(1)}_{\ell L} {\cal E}^{(0)}_{\ell} +{\cal D}^{(0)}_{\ell L} {\cal E}^{(1)}_{\ell} \right ];
\label{eq:b1}\\
&& K_L^{B,2} =  A_2\, {1 \over \Xi_{L}}\sum_\ell {\myC_\ell^{\phi\phi}\over \Xi_{\ell}} 
\left [ \Pi^2_{\ell}{\cal D}^{(0)}_{\ell L} {\cal E}^{(0)}_{\ell}- {\cal D}^{(1)}_{\ell L} {\cal E}^{(1)}_\ell \right ];
\label{eq:b2}\\
&& K_L^{B,3} = A_3\, {1 \over \Xi_{L}}\sum_\ell {\myC_\ell^{\phi\phi}\over \Xi_{\ell}} 
\left [( \Pi_{\ell}{\cal D}^{(0)}_{\ell L} - {\cal D}^{(1)}_{\ell L})( \Pi_{\ell}{\cal E}^{(0)}_{\ell}- {\cal E}^{(1)}_\ell) \right ].
\label{eq:b3}
\een
The amplitudes $A_i$ are same as the ones for the corresponding two-to-two spectra introduced before.
We have introduced the following quantities to simplify our notation:
\ben
&& {\cal D}^{(0)}_{\ell\,L} \equiv \sum_{\ell'} f_{\ell' L\ell}I_{\ell'L \ell}; \quad
{\cal D}^{(1)}_{\ell\,L} \equiv \sum_{\ell'} (\Pi_{\ell}+\Pi_{\ell'}) f_{\ell' L\ell}I_{\ell' L\ell}.
\een
Note that the ${\cal D}$ matrices are not symmetric in their indices.
The quantities ${\cal E}^{(0)}_{L}$ and ${\cal E}^{(1)}_{L}$ defined in Eq.(\ref{eq:E_matrix}) can now be expressed in terms of 
${\cal D}^{(0)}_{\ell L}$ and ${\cal D}^{(1)}_{\ell L}$:
\ben
&& {\cal E}^{(0)}_{L} \equiv \sum_{\ell} {\cal D}^{(0)}_{\ell L}; \quad {\cal E}^{(1)}_{L} \equiv \sum_{\ell} {\cal D}^{(1)}_{\ell L}.
\label{eq:def_D&E}
\een
From Eq.(\ref{eq:b0})-Eq.(\ref{eq:b3}), notice that the three-to-one estimator, unlike the two-to-two estimators, cannot be written
in terms of the lensing power spectrum times an $\ell$-dependent normalisation factor. Instead, it involves a convolution, encapsulated in
the ${\cal D}$ matrices.
For point sources we have:
\ben
&& {\cal D}^{{\rm ps}, (0)}_{\ell L} \equiv \sum_{\ell'} I^2_{\ell' L\ell}; \quad
{\cal D}^{{\rm ps}, (1)}_{\ell L} \equiv \sum_{\ell'} (\Pi_{\ell}+\Pi_{\ell'}) I^2_{\ell'L\ell}.
\een
The covariances of the first two of these estimators can be computed using similar technique as before:
\ben
&& \la \delta K^{B,1}_{\ell} \delta K^{B,1}_{\ell^{\prime}} \ra = \delta_{\ell\ell'}\; 6 {1 \over \Sigma^2_\ell}{\cal C}_{\ell}{ \cal Q}_{\ell} + 
{\cal C}_{\ell}{\cal C}_{\ell'} {1 \over \Xi_{\ell}}{1 \over \Xi_{\ell'}} {\cal R}_{\ell}{\cal R}_{\ell'}; \\
&& \la \delta K^{B,2}_{\ell} \delta K^{B,2}_{\ell^{\prime}} \ra = \delta_{\ell\ell'}\; 6 {\Pi^2_{\ell} \over \Sigma^2_\ell}\; {\cal C}_{\ell}{\cal Q}_{\ell} + 
{\cal C}_{\ell}{\cal C}_{\ell'}{{\Pi}_{\ell} \over \Xi_{\ell}}{{\Pi}_{\ell'}\over \Xi_{\ell'}} {\cal R}_{\ell}{\cal R}_{\ell'}.
\een
The new quantities we have introduced are
\ben
&& {\cal Q}_{\ell}= \sum_{\ell_i=2}\sum_{L}{\cal C}_{\ell_1}{\cal C}_{\ell_2}{\cal C}_{\ell_3} I^2_{\ell_1\ell_2L}I^2_{L\ell_3\ell}\\
&& {\cal R}_{\ell}= \sum_{\ell_2}\sum_L {\cal C}_{\ell_2} I^2_{L\ell_2\ell}.
\een
We have assumed the absence of any parity-violating physics, and ignore the $L=0$, as discussed before.
The results are valid for all-sky coverage. The two-to-two estimators and their three-to-one counterparts
are decorrelated in the Gaussian limit as the leading order terms take contribution from odd-order multispectra.  
\section{Recovery of the Generalised Kurtosis Parameters}
\label{sec:appendB}
The generalised kurtosis can be recovered using either the two-to-two 
or three-to-one kurt-spectra by using the Eq.(\ref{eq:one_pt}), and are given by the following expressions:
\ben
&& K^{(0)} = \sum_L {{\cal C}^{\phi\phi}_{L} \over \Xi_L} {{\cal E}_L^{(0)}}^2; \\
&& K^{(1)}= 2\, \sum_L {{\cal C}^{\phi\phi}_{L} \over \Xi_L} {\cal E}_{L}^{(0)}{\cal E}_{L}^{(1)};\\
&& K^{(2)} = \sum_L {{\cal C}^{\phi\phi}_{L} \over \Xi_L} \left [ \Pi^2_{L}\, {{\cal E}_{L}^{(0)}}^2 - {{\cal E}_{L}^{(1)}}^2 \right ];\\
&& K^{(3)} = \sum_L {{\cal C}^{\phi\phi}_{L} \over \Xi_L} \left [ \Pi_{L}\, {\cal E}_{L}^{(0)} - {{\cal E}_{L}^{(1)}} \right ]^2.
\een
The quantities ${\cal E}^{(0)}_{L}$ and ${\cal E}^{(1)}_{L}$ are given in Eq.(\ref{eq:E_matrix}).
To compute corresponding estimates for the point sources we have to replace ${\cal C}^{\phi\phi}_{L}$ by the amplitude $t_{\rm ps}$
and ${\cal E}_{L}^{(i)}$ with their point-source analogues ${\cal E}_{L}^{\rm ps,(i)}$ defined in Eq.(\ref{eq:E_matrix_ps}).  
\section{Kurt-Spectra as Sub-Optimal Estimators for Lensing Reconstruction}
\label{sec:appendC}
The kurt-spectra  $K_{\ell}^{A,i}$, introduced in this paper in Eq.(\ref{eq:k0})-Eq.(\ref{eq:k4}), 
are constructed from a combination of cross-spectra such as 
$K_\ell^{\Theta^2,\;\Theta^2}$, $K_\ell^{\Theta^2,\Theta\nabla^2\Theta}$, or in general $K_{\ell}^{\Psi_\alpha,\Psi_\beta}$, where $\Psi_{\alpha}$ and $\Psi_{\beta}$ are chosen form $\{ \Theta^2, \Theta\nabla^2\Theta,\nabla\Theta\cdot\nabla\Theta \}$.
The multipole expansion of the derived temperature maps $\Psi_{\alpha}$ are a set of quadratic statistics that can be used as sub-optimal
estimators for reconstruction of lensing potential $\phi$ as we will see below.

The harmonic coefficients of the lensing potential $\phi_{\ell m}$ in general can be 
expressed in terms of such quadratic combination $\Psi$ with a suitable $\ell$ normalization
through a convolution which depends on the weight function $g_{\ell_1\ell_2\ell}$:
\ben
&& \hat\phi_{\ell m} = {1 \over \Psi_\ell}\hat\Psi_{\ell m}; \quad\quad \Psi_{\ell} = \sum_{\ell_1\ell_2} g^{\psi}_{\ell_1\ell\ell_2} f_{\ell_1\ell_2\ell}.\\
&& \hat\phi_{\ell m} =  {1 \over \Psi_\ell} \sum_{\ell_1m_1}\sum_{\ell_2m_2}  (-1)^m  g^{\Psi}_{\ell_1\ell\ell_2}\left ( \begin{array}{ c c c }
     \ell_1 & \ell_2 & \ell \\
     m_1 & m_2 & -m
  \end{array} \right) \hat\Theta_{\ell_1m_1}\hat\Theta_{\ell_2m_2}.
\een 
For a specific choice of $\Psi$ the resulting weights are listed below:
\ben
g_{\ell_1\ell_2\ell}^{\Theta^2} = I_{\ell_1\ell_2\ell}; \quad
g_{\ell_1\ell_2\ell}^{\Theta\nabla^2\Theta}= -\Pi_{\ell_1}I_{\ell_1\ell_2\ell}; \quad
g_{\ell_1\ell_2\ell}^{\nabla\Theta\cdot\nabla\Theta}= {1 \over 2}I_{\ell_1\ell_2\ell}\Lambda_{\ell_1\ell\ell_2}.
\een
This result is completely generic and does not depend on a specific choice of the {\em weighting function} $g_{\ell_1\ell_2\ell}$, although we have approximated $T=P$. Additional terms will contribute to bias and can be removed for any practical application.

Reconstruction of individual harmonics $\phi_{lm}$ is expected to be noise-dominated so the reconstruction is typically
carried out for the power-spectrum $\myC_l^{\phi\phi}$ of the lensing potential $\phi$. 
Thus we can construct a series of estimator for $\myC_{\ell}^{\phi\phi}$ using the two-to-two
estimators. 
\be
\hat \myC_{\ell}^{(\alpha,\beta),\phi\phi} = {1 \over \Psi_\ell^{(\alpha)}\Psi_{\ell}^{(\beta)}}\hat K_{\ell}^{(\alpha,\beta)}
\ee
The other set of estimators that we have studied i.e. $K_l^{B,(i)}$ involves a convolution of $C_{\ell}^{\phi\phi}$ and temperature power spectrum $\bar\myC_l$.
Though these estimators can not be used directly for reconstruction of $\myC_l^{\phi}$ they
can be used for cross-validation of results obtained using $K_l^{A,(i)}$. 

For the optimal estimator presented in Ref.\cite{huOka02} the weight function $g_{\ell_1\ell_2\ell}$ takes the
following form: $g_{\ell_1\ell_2L} = (f_{\ell_1\ell\ell_2}/\bar\myC_{\ell_1}\bar\myC_{\ell_2})$.
Though primarily designed to analyse the morphological properties, they can also work as
sub-optimal estimators for lensing reconstruction and are faster than their optimal counterpart, as 
they can be implemented using the pseudo-$\myC_{\ell}$ approach described in \textsection\ref{sec:estim}.
 
\end{document}